\documentclass[a4paper,fleqn,usenatbib]{mnras}

\usepackage[T1]{fontenc}
\usepackage{ae,aecompl}
\usepackage{epsfig}
\usepackage{epstopdf}

\usepackage{graphicx}	% Including figure files
\usepackage{amsmath}	% Advanced maths commands
\usepackage{amssymb}	% Extra maths symbols
\usepackage{float}
\usepackage{caption}
\usepackage{subcaption}
\usepackage{color}
\usepackage{enumitem}

\usepackage{epstopdf}
\usepackage{float}
\restylefloat{table}
\usepackage{amsmath}
\usepackage{mathtools}
\newcommand{\be}{\begin{equation}}  
\newcommand{\ee}{\end{equation}}
\newcommand{\ba}{\begin{eqnarray*}}\newcommand{\ea}{\end{eqnarray*}}
\usepackage{siunitx}    % for numbers with units
\DeclareSIUnit{\degree}{deg} % we want degrees written with 'deg' and
\DeclareSIUnit{\arcmin}{arcmin} % we want degrees written  'arcmin' and
\DeclareSIUnit{\arcsec}{arcsec} % we want arcsec written  'arcsec' and
\definecolor{purple}{RGB}{76, 0,153}

\newcommand{\am}[1]{\textcolor{black}{ #1}}
\title[Combining weak lensing and LSS]{KiDS$+$2dFLenS$+$GAMA: Testing the cosmological model with the $E_{\rm G}$ statistic}
\author[A. Amon et al.]
{A. Amon$^1$\thanks{Email: aamon@roe.ac.uk}, 
C. Blake$^2$,
C. Heymans$^1$, 
C. D. Leonard$^3$,
M. Asgari$^1$,
M. Bilicki$^{4,5}$,
\newauthor
A. Choi$^{6}$,
T. Erben$^7$,
K. Glazebrook$^{2}$,
J. Harnois-D\'eraps$^{1}$,
H. Hildebrandt$^7$,
\newauthor
H. Hoekstra$^4$,
B. Joachimi$^8$,
S. Joudaki$^{9}$,
K. Kuijken$^4$,
C. Lidman$^{10}$,
\newauthor
J. Loveday$^{11}$,
D. Parkinson$^{12,13}$,
E. A. Valentijn$^{14}$
and
C. Wolf$^{15}$
\\ 
$^1$Institute for Astronomy, University of Edinburgh, Royal Observatory, Blackford Hill, Edinburgh EH9 3HJ, UK\\
$^2$Centre for Astrophysics \& Supercomputing, Swinburne University of Technology, PO Box 218, Hawthorn, VIC 3122, Australia\\
$^3$McWilliams Center for Cosmology, Carnegie Mellon University, 5000 Forbes Ave., Pittsburgh, PA 15213\\
$^4$Leiden Observatory, Leiden University, P.O.Box 9513, 2300RA Leiden, The Netherlands\\
$^5$National Centre for Nuclear Research, Astrophysics Division, P.O. Box 447, PL-90-950 Lodz, Poland\\
$^{6}$Center for Cosmology and AstroParticle Physics, The Ohio State University, 191 West Woodruff Avenue, Columbus, OH 43210, USA\\
$^7$Argelander-Institut f\"ur Astronomie, Auf dem H\"ugel 71, 53121 Bonn, Germany\\
$^8$Department of Physics and Astronomy, University College London, Gower Street, London WC1E 6BT, UK\\
$^9$Department of Physics, University of Oxford, Denys Wilkinson Building, Keble Road, Oxford OX1 3RH, UK\\
$^{10}$Australian Astronomical Observatory, PO Box 915, North Ryde, NSW 1670, Australia\\
$^{11}$Astronomy Centre, University of Sussex, Falmer, Brighton BN1 9QH, UK\\
$^{12}$School of Mathematics and Physics, University of Queensland, Brisbane, QLD 4072, Australia\\
$^{13}$Korea Astronomy and Space Science Institute, Daejeon 34055, Korea\\
$^{14}$Kapteyn Astronomical Institute, University of Groningen, 9700AD Groningen, the Netherlands\\
$^{15}$Research School of Astronomy and Astrophysics, Australian National University, Canberra, ACT 2611, Australia\\
\vspace{-0.5cm}  %necessary so a LaTeX crash doesn't occur...
}

\date{Accepted XXX. Received YYY; in original form ZZZ}

\pubyear{2017}

\begin{document}
\label{firstpage}
\pagerange{\pageref{firstpage}--\pageref{lastpage}}
\maketitle

% Abstract of the paper
\begin{abstract}
We present a new measurement of $E_{\rm G}$, which combines measurements of weak gravitational lensing, \am{real-space} galaxy clustering and redshift space distortions. This statistic was proposed as a consistency test of General Relativity (GR) that is insensitive to linear, deterministic galaxy bias and the matter clustering amplitude. We combine deep imaging data from KiDS with overlapping spectroscopy from 2dFLenS, BOSS DR12 and GAMA and find $E_{\rm G}(\overline{z}=0.267)=0.43 \pm 0.13$ (GAMA), $E_{\rm G}(\overline{z}=0.305)=0.27 \pm 0.08$ (LOWZ+2dFLOZ) and $E_{\rm G}(\overline{z}=0.554)=0.26 \pm 0.07$ (CMASS+2dFHIZ).
%showing roughly a 2$\sigma$ deviation from the GR predictions with a Planck 2015 cosmology. 
We demonstrate that the existing tension in the value of the matter density parameter hinders the robustness of this statistic as solely a test of GR. We find that our $E_{\rm G}$ measurements, as well as existing ones in the literature, favour a lower matter density cosmology than the Cosmic Microwave Background. For a flat $\Lambda$CDM Universe \am{and assuming GR}, we find $\Omega_{\rm m}(z=0)=0.25\pm0.03$. With this paper we publicly release the 2dFLenS dataset at: \url{http://2dflens.swin.edu.au}.
\end{abstract}

\begin{keywords}
gravitational lensing: weak -- surveys, cosmology: observations, large-scale structure of Universe  

\vspace{0.4cm}
\end{keywords}

%%%%%%%%%%%%%%%%%%%%%%%%%%%%%%%%%%%%%%%%%%%%%%%%%%

%%%%%%%%%%%%%%%%% BODY OF PAPER %%%%%%%%%%%%%%%%%%

\section{Introduction} %\input introduction.tex
Many observations reveal that within the Friedmann-Robertson-Walker (FRW) framework, the Universe is undergoing a late-time, accelerated expansion, which is driven by some unknown `dark energy' \citep[see for example][]{Copeland2006}. While a vacuum energy is the simplest and most widely accepted model of dark energy, there exists an enormous discrepancy between its theoretical and observed value \citep{Weinberg1989}. To address this problem, a wide range of alternative models have been proposed including those where gravity behaves differently on large cosmological scales from the framework laid down by Einstein's General Relativity (GR). As an understanding of the nature of this dark energy phenomenon still evades scientists, it is imperative that current cosmological surveys conduct observations to test for such departures on cosmological scales \citep{Weinberg2012}.

The perturbed FRW spacetime metric may be completely defined in terms of the Bardeen potentials \citep{Bardeen1980}, namely the Newtonian potential, $\Psi$, which along with density perturbations drives the structure formation of the Universe, and the curvature potential, $\Phi$, as well as an expansion scale factor for the Universe, $a(t)$, as,
\begin{equation}
\label{eqn:bardeen}
\centering
{\rm d}s^2 = -c^2{\rm d}t^2(1+2\Psi)+ a(t)^2 {\rm d}\mathbf{x}^2 (1-2\Phi) \, ,
\end{equation}
where {\bf x} represents the spatial elements of the metric. While cosmological probes by themselves can be subject to model degeneracies and systematic biases, a combination of probes, specifically using imaging and spectroscopic surveys, can test for departures from GR \citep[see for example][]{Zhang2007, JainZhang2008}. The Bardeen potentials are equal in the absence of anisotropic stress, as in the case of GR. This is not necessarily the case in modified gravity theories \citep{PogasianSilvesti}, although recent gravitational wave measurements have set tight constraints on these scenarios \citep{Lom2016, Lom2017, Amendola2017, Baker2017, Creminelli2017, Ezquiaga2017, Sakstein2017}.

Weak gravitational lensing, a statistical quantification of the deflection of light by over-densities in the Universe, has proven itself to be a powerful cosmological probe \citep[see for example][]{Heymans2013,Hildebrandt/etal:2017,  Troxel2017}. This measurement is sensitive to the curvature potential, $\nabla^2(\Psi-\Phi)$, because relativistic particles collect equal contributions from the two potentials as they traverse equal quantities of space and time. One particular observable, galaxy-galaxy lensing, measures the deflection of light due to the gravitational potential of a set of foreground lens galaxies, rather than the large-scale structure as a whole \citep{HoekstraYeeGladder2004, Mandelbaum/etal:2005}.

The clustering effect of the non-relativistic peculiar motions of foreground galaxies can be quantified by measuring redshift-space distortions \citep[RSD;][]{Kaiser1987}. The gravity-driven motion produces Doppler shifts in galaxy redshifts that are correlated with each other. As a result, an overall anisotropy is imprinted in the measured redshift-space clustering signal that is a function of the angle to the line-of-sight. This anisotropy is the redshift-space distortion and an accurate measurement of its amplitude probes the growth rate of cosmic structure, $f$. These probes are sensitive only to derivatives of the Newtonian potential, $\nabla^2\Psi$ and as such, in conjunction with the lensing signal due to the foreground lens galaxies,
allows us to isolate the relativistic deflection of light from background galaxies. This
creates a fundamental test of the relationship between $\Psi$ and $\Phi$.

The complementarity between imaging and spectroscopic surveys has been exploited in the examination of the level of concordance of cosmological measurements from combined lensing, clustering and/or redshift-space distortion analyses \citep{vanU2017, Joudaki2017}, compared to cosmic microwave background (CMB) temperature measurements from the Planck satellite \citep{Planck2016}. These combined-probe analyses \citep[see also][]{DES2017} found varying levels of `tension' with the Planck CMB measurements. In this analysis we combine lensing, \am{real-space} clustering and redshift-space distortions measurements to probe the $E_{\rm G}$ statistic \citep{Zhang2007}. The relative amplitude of the observables is used to determine whether GR's predictions hold, assuming a perturbed FRW metric and a defined set of cosmological parameters. . Any deviations on large scales from the GR prediction for $E_{\rm G}$, which is scale-independent will suggest a need for large-scale modifications in gravitational physics.

As a choice is made for the cosmology used to compute a GR prediction for $E_{\rm G}$, this brings into question the use of this statistic to test GR while any uncertainty exists in the values of the cosmological parameters. This is relevant as there exists a current `tension' in the literature between cosmological parameters (specifically $\sigma_8\sqrt{\Omega_{\rm m}/0.3}$) constrained by Planck CMB experiments \citep{Planck2016} and lensing or combined probe analyses. More specifically, \citet{Hildebrandt/etal:2017} and \citet{Joudaki2017} report a $2.3\sigma$ and $2.6\sigma$ discordance with Planck constraints. We investigate whether the deviations we find from a Planck GR prediction are consistent with the expectations given by the existing tension between early Universe and lensing cosmologies. Even with this uncertainty however, the $E_{\rm G}$ statistic still provides a test of the theory of gravity through its scale dependence. We conduct this test, while investigating the possibility of this effect's degeneracy with scale-dependent bias.

The power of combined-probe analyses was investigated by, for example \citet{Zhao2009, Cai2011, Joudaki2012} and later applied to data \citep{Tereno2011, Simpson2013, ZhaoBacon2015, Planck2016, Joudaki2017}. 
In this paper we extend the original $E_{\rm G}$ measurement performed by \citet{Reyes2010} in redshift and scale, using the on-going large-scale, deep imaging Kilo-Degree Survey \citep[KiDS;][]{Kuijken/etal:2015} in tandem with the overlapping spectroscopic 2-degree Field Lensing Survey \citep[2dFLenS;][]{Blake/etal:2016}, the Baryon Oscillation Spectroscopic Survey \citep[BOSS;][]{Dawson/etal:2013} and the Galaxy and Mass Assembly survey \citep[GAMA;][]{Driver/etal:2011}. With the combination of these data, we extend the statistic  to $\sim50 \, h^{-1}$Mpc in three redshift ranges. \citet{Alam2016}, \citet{Blake2016eg},  and \citet{sdlt2016} previously probed the same high-redshift range and the latter two cases find some tension between their measurements compared to a Planck cosmology. \citet{Pullen2016} measured $E_{\rm G}$ with a modified version of the statistic that incorporates CMB lensing and allows them to test larger scales, finding a 2.6$\sigma$ deviation from a GR prediction, also computed with a Planck cosmology. A number of possible theoretical systematics, as well as predictions for $E_{\rm G}$ in phenomenological modified gravity scenarios are discussed in \citet{Leonard2015b}. 
 
This paper is structured as follows. Section \ref{sec:th} describes the underlying theory of our observables. An outline of the various datasets and simulations involved in the analysis is given in Section \ref{sec:data}. In Section \ref{sec:measurements} we present the different components of the $E_{\rm G}$ statistic and detail how those measurements were conducted, while in Section \ref{sec:results}, we provide our main $E_{\rm G}$ measurement in comparison to existing measurements, as well as to models using different cosmologies and with alternative theories of gravity. We summarise the outcomes of this study and provide an outlook in Section \ref{sec:conc}. 

%%%%%%%%%%%%%%%%%%%%%%%%%%%%%%%%%%%%%%%%%%%%%%%%%%%%%%%

\section{Theory} \label{sec:th} %\input theory.tex
\subsection{Differential surface density}  
Galaxy-galaxy lensing can be mathematically expressed in terms of the cross-correlation of a galaxy overdensity, $\delta_{\rm g}$, and the underlying matter density field, $\delta_{\rm m}$, given at a fixed redshift by $\xi_{\rm gm}(\bf{r})=\langle \delta_{\rm g}(\bf{x})\delta_{\rm m}(\bf{x}+\bf{r})\rangle_{\bf{x}}$.
In order to measure the lensing galaxy-matter cross-correlation function, $\xi_{\rm gm}$, one can first determine the comoving projected surface mass density, $\Sigma_{\rm com}$, around a foreground lens at redshift $z_{\rm l}$,  using a background galaxy at redshift $z_{\rm s}$ and at a comoving projected radial separation from the lens, $R$. This is given as, 
\begin{equation}
\label{eqn:deltasigdef}
\centering
\Sigma_{\rm com}(R)= \overline{\rho_{\rm m}} \int_{0}^{\chi(z_{\rm s})}  \ \xi_{\rm gm} \big( \sqrt{R^2+[\chi-\chi(z_{\rm l})]^2} \big) \ \rm{d}\chi \, ,
\end{equation}
where $\overline{\rho_{\rm m}}$ is the mean matter density of the Universe, $\chi$ is the comoving line-of-sight separation and ${\chi(z_{\rm l})}$, ${\chi(z_{\rm s})}$ are the comoving line-of-sight distances to the lens and source galaxy, respectively. The shear is sensitive to the density contrast, therefore, it is a measure of the excess or differential surface mass density, $\Delta \Sigma_{\rm com}(R)$ \citep{Mandelbaum/etal:2005}. This is defined in terms of $\Sigma_{\rm com}(R)$ as,
\be
\Delta \Sigma_{\rm com}(R)=\overline{\Sigma}_{\rm com}(\leq R) - \Sigma_{\rm com}(R) \, ,
\ee
where the average projected mass density within a circle is,
\be
\label{eqn:apmd}
\overline{\Sigma}_{\rm com}(\leq R)=\frac{2}{R^2} \int^R_0 \Sigma_{\rm com}(R') R' \rm{d}R' \, .
\ee
For a sufficiently narrow lens distribution (such that it may be approximated as a Dirac Delta function at $z_{\rm l}$), in the context of General Relativity, the physical differential surface mass density of the lens is related to the tangential shear, $\gamma_{\rm t}$, of background galaxies as
\begin{equation}
\centering
\Delta\Sigma_{\rm phys}(R) = \gamma_{\rm t}(R)\Sigma_{\rm c, phys} \, .
\label{eqn:ds}
\end{equation}
where $\Sigma_{\rm c, phys}$ is the critical surface mass density. This is defined as, 
\begin{equation}
\label{eqn:scviola}
\centering
\Sigma_{\rm c, phys}= \frac{c^2}{4\pi G} \frac{D(z_{\rm s})}{D(z_{\rm l}) D(z_{\rm l},z_{\rm s})} \, ,
\end{equation}
where $D(z_{\rm s})$, $D(z_{\rm l})$, $D(z_{\rm l},z_{\rm s})$ are the angular diameter distances to the source, to the lens and the angular diameter distance between the source and lens, respectively, $G$ is the gravitational constant and $c$ is the speed of light.
The surface mass density or the convergence, $\kappa$, can be expressed as the ratio of the physical projected and critical surface mass densities, $\Sigma_{\rm phys}$ and $\Sigma_{\rm c, phys}$, respectively, or the equivalent in comoving units, as
\begin{equation}
\centering
\kappa= \frac{\Sigma_{\rm phys}}{\Sigma_{\rm c, phys}}= \frac{\Sigma_{\rm com}}{\Sigma_{\rm c,com}} \, ,
\end{equation}
where the comoving and physical critical surface mass densities\footnote{We note that $\Sigma_{\rm c,com}$ is denoted as $\Sigma_{\rm c}$ in, for example, \citet{Mandelbaum/etal:2005, Leauthaud2016, Blake2016eg, Miyatake:/etal2015, Singh2016}, whereas $\Sigma_{\rm c,phys}$  is denoted as $\Sigma_{\rm c}$ in, for example, \citet{vanUitert2011,Viola2015, Prat2017}.} 
for a lens at redshift $z_{\rm l}$ are related by,
\begin{equation}
\label{eqn:scconvert}
\centering
\Sigma_{\rm c, com}= \frac{\Sigma_{\rm c, phys}}{(1+z_{\rm l})^2} \, .
\end{equation}
The cross-correlation of the lens galaxies and the underlying mass, $\xi_{\rm gm}(r)$, that appears in the definition of the differential surface mass density given by equation~\ref{eqn:deltasigdef}, depends on the way that the lens galaxies trace their matter field. This is known as the `galaxy bias', $b$, and it can be stochastic, non-linear and scale-dependent on small scales \citep{Dekel1999}. However, on linear scales, the galaxy overdensity is expected to be related to the matter overdensity as,
\begin{equation}
\centering
\delta_{\rm g}(\boldsymbol{x})=b \  \delta_{\rm m}(\boldsymbol{x}) \, ,
\end{equation}
so that 
\begin{equation}
\centering
\xi_{\rm gm}(r)=b \ \xi_{\rm mm}(r) \, ,
\end{equation}
where $\xi_{\rm mm}(r)$ is the matter autocorrelation function, which can be derived from the cosmological model \citep{Kaiser1984}.

\subsection{Galaxy Clustering: Redshift-Space Distortions} 
An observed redshift has a contribution from the expansion of the Universe, known as the cosmological redshift, and another from the peculiar velocity.
Measurements sensitive to the peculiar velocities of galaxies are a particularly useful tool for testing gravitational physics.  Peculiar velocities are simply deviations in the motion of galaxies from the Hubble flow due to the gravitational attraction of objects to surrounding structures. 

The two-point statistics of the correlated positions of galaxies in redshift-space are a powerful tool for testing GR growth predictions \citep{Guzzo2008}. Large-scale clustering in real space is isotropic. However, redshift-space distortion introduces a directional dependence such that the redshift-space power spectrum under the assumption of linear theory is
\begin{equation}
\centering
P_{\rm gg} (k,\eta)= b^2(1+ \beta \eta^2)^2 P_{\rm mm}(k)\, ,
\end{equation}
where $P_{\rm mm}$ is the real space matter power spectrum and $\eta$ is the cosine of the angle of the Fourier mode to the line-of-sight \citep{Hamilton1993}.
The factor $\beta$ is introduced as a redshift-space distortion parameter which governs the anisotropy of the clustering amplitude on the angle to the line-of-sight. 
This factor is defined as,
\begin{equation}
\centering
\label{eqn:beta}
\beta \equiv \frac{f(z)}{b(z)} \, ,
\end{equation}
where $f(z)$ is the growth rate of structure. It can be expressed in terms of the growth factor $D_+(a)$ at a particular cosmic scale factor, $a$, defined in terms of the amplitude of the growing mode of a matter-density perturbation as $\delta_{\rm m}(a) = D_+(a)\delta_{\rm m}(z=0)$ to give,
\begin{equation}
\centering
f(z)\equiv \frac{{\rm d} \, \ln D_+(a)}{{\rm d} \, \ln a} \, .
\end{equation}
As a function of the matter density parameter, in the absence of anisotropic stress in GR and with a flat Universe,  the growth rate is well-approximated in terms of the matter density parameter at a given redshift, $\Omega_{\rm m}(z)$, as $f(z)\approx \Omega_{\rm m}(z)^{0.55}$ \citep{WangSteinhardt1999, Linder2005}.

\subsection{Galaxy Clustering: Projected Correlation Function} 
Galaxy clustering independent of RSD can be analysed in terms of the projected separation of galaxies on the sky. We call the associated two-point function in real space the `projected correlation function', $w_{\rm p}(R)$, and it is formulated from the integral of the 3D galaxy correlation function, $ \xi_{\rm gg}(R, \Pi)$, along the line of sight as,
\begin{equation}
\label{eqn:wp}
\centering
w_{\rm p}(R)= \int^{+\infty}_{-\infty}\xi_{\rm gg}(R, \Pi ) \ d\Pi \, ,
\end{equation}
where $\Pi$ is the co-moving separation along the line-of-sight.
%Under the further assumption of linear bias, this additional measurement from galaxy clustering allows for the estimation of the galaxy bias and the shape of $\xi_{\rm mm}(r)$ as $\xi_{\rm gg}(r) = b^2 \xi_{\rm mm}(r)$.
%\am{The anisotropic imprint of RSD in galaxy clustering introduces an extra amplitude factor in the relation to the shape of $\xi_{\rm mm}(r)$ as $\xi_{\rm gg}(r) = b^2 \xi_{\rm mm}(r)$, which contaminates inferences about the galaxy bias.}

\subsection{Suppressing Small Scale Systematics} 
It is evident that the differential surface density of matter, defined in equation~\ref{eqn:apmd}, includes a range of smaller scales from zero to $R$. However, the cross-correlation coefficient between the matter and the galaxy fluctuations is a complicated function at scales within the halo virial radius \citep{Cacciatio2012} and furthermore, lensing systematics can dominate on small scales \citep{Mandelbaum2009}. Thus, in order to reduce the measurement's systematic uncertainty \am{and to circumvent the difficulty in modelling the most non-linear scales,} its sensitivity to small-scale information should be suppressed. This is achieved through a statistic, the comoving annular differential surface density, proposed by \citet{Mandelbaum2009} as,
\begin{equation}
\label{eqn:upsgm}
\centering
\Upsilon_{\rm gm}(R,R_0)= \Delta \Sigma_{\rm com}(R)- \frac{R_0^2}{R^2}\Delta \Sigma_{\rm com}(R_0) \, ,
\end{equation}
where $R_0$ is the small-scale limit below which information is erased. The minimum length scale is chosen to be large enough to reduce the dominant systematic effects, but small enough to maintain a high-signal-to-noise ratio in the galaxy-lens correlation measurement. 
A similar statistic is formulated in order to remove the small-scale contribution to the galaxy auto-correlation,
\begin{equation}
\centering
\label{eqn:upsgg}
\Upsilon_{\rm gg}(R,R_0)= \rho_{\rm c} \Big[ \frac{2}{R^2} \int_{R_0}^{R}  R' w_{\rm p}(R')dR'-w_{\rm p}(R)+\frac{R_0^2}{R^2}w_{\rm p}(R_0)\Big] \, ,
\end{equation}
where $\rho_{\rm c}$ is the critical density. 

We note that an alternative method to remove small-scale systematics was introduced in \citet{Buddendiek2016}, where they generalised the $\Upsilon$ formalism from \citet{Baldauf2010} by an expansion of the galaxy-galaxy and galaxy-matter correlation functions using a complete set of orthogonal and compensated filter functions. This is inspired by COSEBIs for the case of cosmic shear analysis defined in \citet{SchneiderEiflerKrause2010} \citep[see][for an application to data]{Asgari2017}.

\subsection{The $E_{\rm G}$ Statistic} \label{sec:EG}
The $E_{\rm G}$ statistic, as proposed by \citet{Reyes2010}, is defined as a combination of the annular statistics $\Upsilon_{\rm gm}(R,R_0)$ and $\Upsilon_{\rm gg}(R,R_0)$ with the RSD parameter, $\beta$ (equations~\ref{eqn:upsgm}, \ref{eqn:upsgg}, \ref{eqn:beta}, respectively) as,
\begin{equation}
\centering
E_{\rm G}(R)=\frac{1}{\beta} \frac{\Upsilon_{\rm gm}(R,R_0)}{\Upsilon_{\rm gg}(R,R_0)} \, ,
\label{eqn:eg}
\end{equation}
where all of the statistics are measured for a particular lens galaxy sample. This measurement depends on the redshift of the lens galaxy sample, $z_{\rm l}$, but we omit this for clarity. In this combination, the contribution of a linear galaxy bias, as well as the shape of the matter clustering in the measurements approximately cancel.

The gravitational statistic was initially theorised by \cite{Zhang2007}, who proposed it as an estimator of the form,
\begin{equation}
\tilde{E}_{\rm G}(l) = \frac{C_{\rm{g} \kappa}(l)}{3 H_0^2 a^{-1}(z) C_{\rm{g} v} (l)}\, ,
\label{eqn:EGFourier}
\end{equation}
where $C_{\rm{g} \kappa}$ is the projected cross-power spectrum of source galaxy convergence with lens galaxy positions, $l$ is the amplitude of the on-sky Fourier-space variable conjugate to projected radius, $H_0$ is the Hubble parameter today and $C_{\rm{g} v} (l)$ is a projected version of the cross-power spectrum of lens galaxy positions and velocities. The theoretical expectation value of this statistic, averaged over $l$ \am{and under linear regime}, is predicted to take the value,
\begin{equation}
\centering
E_{\rm G}(z)=\frac{ \nabla^2[\Psi(z)+\Phi(z)]}{3H_0^2 a^{-1}(z) f(z) \delta_{\rm m}(z)}\, ,
\end{equation} 
where $\delta_{\rm m}$ is the matter field overdensity and $\Psi(z)$ and $\Phi(z)$ are the Bardeen potentials from equation~\ref{eqn:bardeen}. We do not indicate the $k$-dependences of $\Psi, \Phi$ and $\delta_{\rm m}$ as in linear regime, these cancel, thereby rendering $E_{\rm G}(z)$ independent of $k$. Invoking mass-energy conservation in a standard FRW Universe, and under the assumption that we are in the linear regime, results in $\Psi=\Phi$ and
\begin{equation}
\centering
\nabla^2\Phi=\nabla^2\Psi=\frac{3}{2}\Omega_{\rm m}(z=0)H_0^2a(z)^{-1}\delta_{\rm m}(z) \, .
\end{equation}
\citet{Zhang2007} showed that this can be reduced to a value of $E_{\rm G}(z)$ which is a function of the matter density parameter valued today, $\Omega_{\rm m}(z=0)$, and the growth rate of structure, $f(z)$, that is independent of the comoving scale $R$ and defined to be
\begin{equation}
\centering
E_{\rm G}(z)=\frac{\Omega_{\rm m}(z=0)}{f(z)} \, .
\label{EG_theory}
\end{equation}
Here the dependence of this statistic on an underlying cosmology is evident.
As the prediction from GR is scale-independent \am{in the linear regime}, it is useful to compress the observable defined in equation~\ref{eqn:eg} to a scale-independent measurement at the effective redshift of the lens sample, $E_{\rm G}(\hat{z})=\langle E_{\rm G}(R) \rangle$.

The elegance of the statistic proposed by \citet{Zhang2007} is that it is constructed to be independent of the poorly-constrained galaxy bias factor, $b$, given that on large scales, linear theory applies. However, measuring $E_{\rm G}$ following equation~\ref{eqn:EGFourier} requires a Fourier space treatment of probes which are typically analysed in real space, as well as a measurement of the cross-spectra of galaxy positions with convergence and velocities, which are in practice challenging to determine directly. The real-space statistic of equation~\ref{eqn:eg} is hence the more convenient estimator and the one we employ in this paper. \citet{Leonard2015} showed that in the case of linear bias, a flat cosmology and in GR, $E_{\rm G}=\tilde{E}_{\rm G}$. It is worth noting, however, that in real space we lose the ability to cleanly restrict the measurement to the linear regime. Therefore, it is less clear at which scales $E_{\rm G}$ remains independent of galaxy bias. As shown in \cite{Alam2016} using N-body simulations, this effect is expected to be at most of order 8 percent at 6$h^{-1}{\rm Mpc}$ for Luminous Red Galaxies, and therefore is unlikely to affect our results significantly. We explore the effect of galaxy bias in the measurement in Section~\ref{sec:results}.

\subsection{Modifications to gravity} \label{sec:MG}
$E_{\rm G}$ is designed, in theory, as a model-independent probe of gravity, such that one does not need to test any one particular theory of gravity or define a specific form for the deviations from General Relativity. However, in order to compare this measurement to other analyses, we consider a phenomenological parameterisation of deviations from GR in a quasistatic regime. This parameterisation is valid under the approximation that within the range of scales accessible to our data, any time derivatives of new gravitational degrees of freedom are set to zero.  This approximation has been shown to hold in most cosmologically-motivated theories of gravity on the range of scales relevant to this measurement \citep{Noller2014, Schmidt2009, Zhao2011, Barreira2013, Li2013}.  In the version of this parameterisation that we employ here, the modifications to gravity are summarised as alterations to the Poisson equation for relativistic and non-relativistic particles as \citep[e.g.][]{Simpson2013},
\begin{equation}
\begin{aligned}
\label{eqn:mg1}
2\nabla^2\Psi(z,k) &= 8\pi G a(z)^2[1 +\mu(z,k)]\rho_{\rm m}\delta_{\rm m}(z,k) \\
\nabla^2 ( \Psi(z,k) + \Phi(z,k)) &= 8\pi G a(z)^2[1 + \Sigma(z,k)]\rho_{\rm m}\delta_{\rm m}(z,k) \, .
\end{aligned}
\end{equation}
We model $\mu$ and $\Sigma$ as small deviations from GR+$\Lambda$CDM, \am{assuming flatness} and following \citet{Ferreira2010, Simpson2013} as,
\begin{equation}
\begin{aligned}
\centering
\Sigma(z) &=\Sigma_0 \frac{\Omega_{\Lambda}(z)}{\Omega_{\Lambda}(z=0)} \\
\mu(z) &=\mu_0 \frac{\Omega_{\Lambda}(z)}{\Omega_{\Lambda}(z=0)} \, ,
\end{aligned}
\end{equation}
where $\mu_0$ and $\Sigma_0$ are the present-day values for the parameters $\mu$ and $\Sigma$ and govern the amplitude of the deviations from GR \am{and $\Omega_{\Lambda}$ is the density parameter for $\Lambda$}. This choice of redshift dependence is selected because in the case in which deviations from General Relativity are fully or partially responsible for the accelerated expansion of the Universe, we would expect $\mu$ and $\Sigma$ to become important at the onset of this acceleration. This form for $\mu$ and $\Sigma$ assumes that any scale-dependence of modifications to GR is sub-dominant to redshift-related effects. Within the regime of validity of the quasistatic approximation, this has been demonstrated to be a valid assumption \citep{Silvestri2013}. We also assume a scale-independent galaxy bias.

Within this scale-independent anzatz for $\mu$ and $\Sigma$ and assuming small deviations from GR, $E_{\rm G}$ is predicted to be given by
\begin{equation}
E_{\rm G}(z) = [1 + \Sigma(z)] \frac{\Omega_{\rm m}(z=0)}{f[z, \mu(z)]}\, ,
\end{equation}
where the dependence of $f(z)$ on the deviation of the Poisson equation from its GR values is given explicitly for clarity in \citet{Baker2014,Leonard2015}. 

\begin{figure*}
\centering
\includegraphics[width=\textwidth]{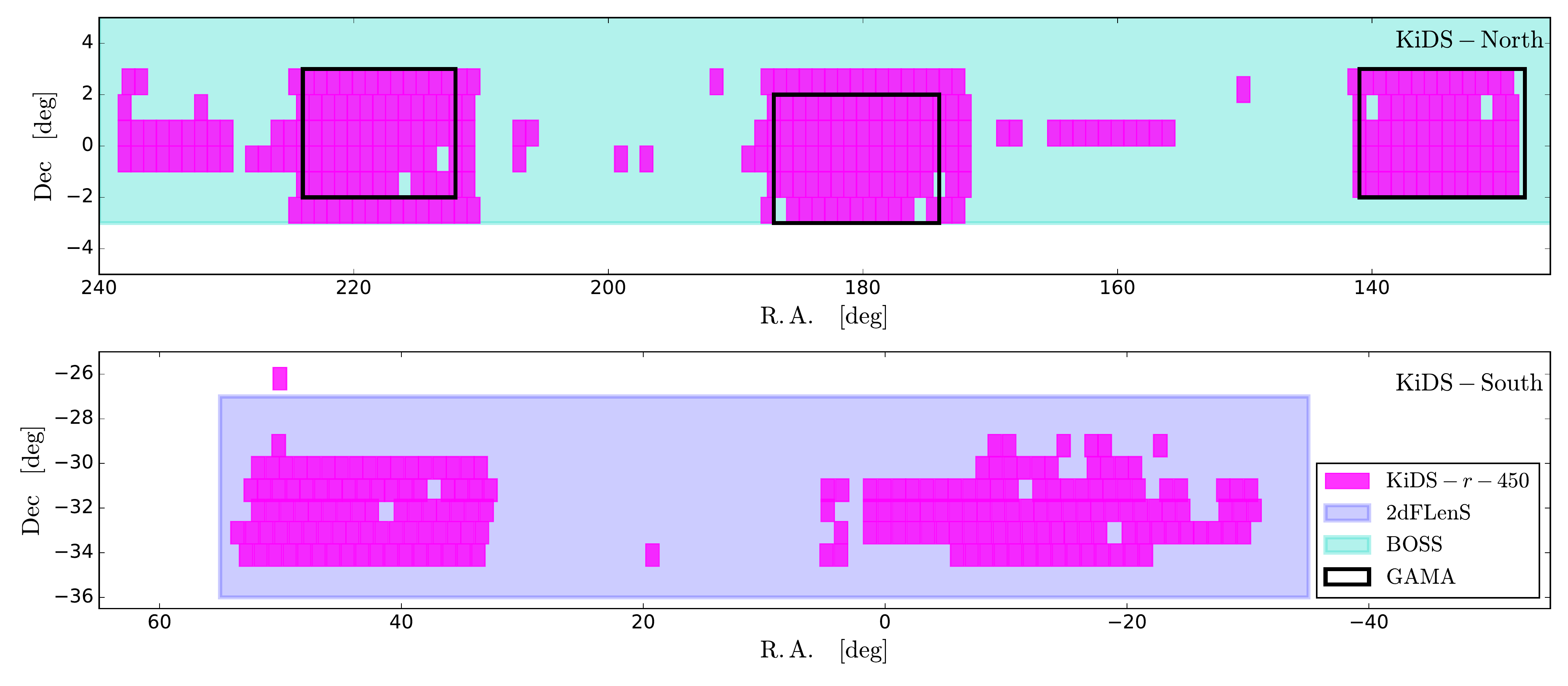}
\caption{\label{fig:map} KiDS-450 survey footprint. Each pink box corresponds to a single KiDS pointing of $\SI{1}{\square\degree}$. The turquoise region indicates the overlapping BOSS coverage and the blue region represents the 2dFLenS area. The black outlined rectangles are the GAMA spectroscopic fields that overlap with the KiDS North field.}
\end{figure*}

%%%%%%%%%%%%%%%%%%%%%%%%%%%%%%%%%%%%%%%%%%%%%%%%%%%%%%%

\section{Data and Simulations} \label{sec:data} %\input dataandsims.tex
\subsection{Kilo Degree Survey (KiDS)} 
\begin{table*}
\caption{ \label{tab:spec} For each spectroscopic survey used in the analysis, this Table quotes the full area used for the clustering analysis, $A_{\rm full}$, the overlapping effective area $A_{\rm eff}$ with the KiDS imaging and the number of lenses in the overlap region of each sample that were used in the lensing analyses. Also quoted are the mean redshift of the spectroscopic sample and the RSD measurements of the $\beta$ parameter, taken from \citet{Blake/etal:2016} for 2dFLenS, \citet{Singh2018} for the BOSS samples and \citet{Blake2013} for the analysis with GAMA. } 
\begin{center}
\begin{tabular}{lcccccc}
 \hline
Spec. sample & $A_{\rm full}$  (deg$^2$) & $A_{\rm eff}$ (deg$^2$) & $N_{\rm lenses}$ & $\overline{z}$ & $\beta$ \\
 \hline
GAMA & 180 & 144 & 33682 & 0.267 & 0.60 $\pm$ 0.09 \\ 
LOWZ & 8337 & 125 & 5656 & 0.309 & 0.41 $\pm$ 0.03 \\
CMASS & 9376 & 222 & 21341 & 0.548 & 0.34 $\pm$ 0.02 \\
2dFLOZ & 731 & 122 & 3014 & 0.300 & 0.49 $\pm$ 0.15 \\ 
2dFHIZ & 731 & 122 & 4662 & 0.560 & 0.26 $\pm$ 0.09 \\ 
 \hline
\end{tabular}
\end{center}
\end{table*}

The Kilo-Degree Survey (KiDS) is a large-scale, tomographic, weak-lensing imaging survey \citep{Kuijken/etal:2015} using the wide-field camera, OmegaCAM, at the VLT Survey Telescope at ESO Paranal Observatory.  It will span \SI{1350}{\square\degree} on completion, in two patches of the sky with the $ugri$ optical filters, as well as 5 infrared bands from the overlapping VISTA Kilo-degree Infrared Galaxy (VIKING) survey \citep{Edge2013}, yielding the first well-matched wide and deep optical and infrared survey for cosmology. The VLT Survey Telescope is optimally designed for lensing with high-quality optics and seeing conditions in the detection $r$-band filter with a median of $<0.7\arcsec$. 

The fiducial KiDS lensing dataset which is used in this analysis, `KiDS-450', is detailed in \citet{Hildebrandt/etal:2017} with the public data release described in \cite{deJong2017}. This dataset has an effective number density of $n_{\rm eff}=8.5$ galaxies arcmin$^{-2}$ with an effective, unmasked area of \SI{360}{\square\degree}. The KiDS-450 footprint is shown in Figure~\ref{fig:map}. Galaxy shapes were measured from the $r$-band data using a self-calibrating version of \textit{lens}fit \citep{miller/etal:2013, fenech-conti/etal:2016} and assigned a lensing weight, $w_{\rm s}$ based on the quality of that galaxy's shape measurement. Utilising a large suite of image simulations, the multiplicative shear bias was deemed to be at the percent level for the entire KiDS ensemble and is accounted for during our cross-correlation measurement. 

The redshift distribution for KiDS galaxies was determined via four different approaches, which were shown to produce consistent results in a cosmic shear analysis \citep{Hildebrandt/etal:2017}. We adopt the preferred method of that analysis, the `weighted direct calibration' (DIR) method, which exploits an overlap with deep spectroscopic fields. Following the work of \citet{Lima2008}, the spectroscopic galaxies are re-weighted such that any incompleteness in their spectroscopic selection functions is removed. A sample of KiDS galaxies is selected using their associated $z_{\rm B}$ value, estimated from the four-band photometry as the peak of the redshift posterior output by the Bayesian photometric redshift BPZ code \citep{Benitez2000}. The true redshift distribution for the KiDS sample is determined by matching these to the re-weighted spectroscopic catalogue. 
The resulting redshift distribution is well-calibrated in the range $0.1 < z_{\rm B} \leq 0.9$.

KiDS has spectroscopic overlap with the Baryon Oscillation Spectroscopic Survey (BOSS) and the Galaxy And Mass Assembly (GAMA) survey in its Northern field and the 2-degree Field Lensing Survey (2dFLenS) in the South. The footprints of the different datasets used in this analysis are shown in Figure~\ref{fig:map} and the effective overlapping areas are quoted in Table~\ref{tab:spec}. 

\subsection{Spectroscopic overlap surveys} 
BOSS is a spectroscopic follow-up of the SDSS imaging survey, which used the Sloan Telescope to obtain redshifts for over a million galaxies spanning $\sim$\SI{10000}{\square\degree}. BOSS used colour and magnitude cuts to select two classes of galaxies: the `LOWZ' sample, which contains Luminous Red Galaxies (LRGs) at $z_{\rm l} < 0.43$, and the `CMASS' sample, which is designed to be approximately stellar-mass limited for $z_{\rm l} > 0.43$.  We used the data catalogues provided by the SDSS 12th Data Release (DR12); full details of these catalogues are given by \cite{Alam/etal:2015}.  Following standard practice, we select objects from the LOWZ and CMASS datasets with $0.15 < z_{\rm l}  < 0.43$ and $0.43 < z_{\rm l}  < 0.7$, respectively, to create homogeneous galaxy samples. In order to correct for the effects of redshift failures, fibre collisions and other known systematics affecting the angular completeness, we use the completeness weights assigned to the BOSS galaxies \citep{Ross2016}, denoted as $w_{\rm l}$. 
The RSD parameters, $\beta$ for LOWZ and CMASS are quoted in Table~\ref{tab:spec} and were drawn from \citet{Singh2018}, who follow the method described in \citet{Alam2015}. This analysis used the monopole and quadrupole moments of the galaxy auto-correlation function, obtained by projecting the redshift space correlation function on the Legendre basis. These multipole moments were fit in each case applying a perturbation theory model, using scales larger than 28$h^{-1}$Mpc and fixing the Alcock-Paczynski parameters. This fitting range excludes the small scales that are used in our clustering and lensing measurements and we therefore assume that the RSD parameters are relatively constant across linear scales. An improvement to future $E_{\rm G}$ measurements can come from better RSD modelling to the small scales. 

2dFLenS is a spectroscopic survey conducted by the Anglo-Australian Telescope with the AAOmega spectrograph, spanning an area of \SI{731}{\square\degree} \citep{Blake/etal:2016}. It is principally located in the KiDS regions, in order to expand the overlap area between galaxy redshift samples and gravitational lensing imaging surveys.  The 2dFLenS spectroscopic dataset contains two main target classes: $\sim$\SI{40000} Luminous Red Galaxies (LRGs) across a range of redshifts $z_{\rm l}< 0.9$, selected by BOSS-inspired colour cuts \citep{Dawson/etal:2013}, as well as a magnitude-limited sample of $\sim$\SI{30000} objects in the range $17<r<19.5$, to assist with direct photometric calibration \citep{wolf/etal:2017, bilicki2017}.  In our study we analyse the 2dFLenS LRG sample, selecting redshift ranges $0.15<z_{\rm l}<0.43$ for `2dFLOZ' and $0.43<z_{\rm l}<0.7$ for `2dFHIZ', mirroring the selection of the BOSS sample.  We refer the reader to \cite{Blake/etal:2016} for a full description of the construction of the 2dFLenS selection function and random catalogues. The RSD parameter was determined by \citet{Blake/etal:2016} from a fit to the multipole power spectra and was found to be $\beta=0.49 \pm 0.15$ and $\beta=0.26 \pm 0.09$ in the low- and high-redshift LRG samples, respectively.  \am{This analysis used a standard model for the redshift-space galaxy power spectrum as a function of the angle of the Fourier wave vector to the line of sight. The model is characterised by three parameters: the galaxy bias factor, $b$, a free parameter for the velocity dispersion, $\sigma_{\rm v} ({\rm km \, s}^{-1})$ which combines the large-scale `Kaiser limit' amplitude correction with a heuristic damping of power on smaller scales that describes a leading-order perturbation theory correction and of course, the RSD parameter, $\beta$. This 3-parameter model fixes the Alcock-Paczynski parameters and is fit to the monopole and quadrupole of both the power spectrum and the correlation function and is found to be relatively insensitive to the fitting range.} We present the 2dFLenS data release in Section~\ref{sec:2df}. 

GAMA is a spectroscopic survey carried out on the Anglo-Australian Telescope with the AAOmega spectrograph. We use the GAMA galaxies from three equatorial regions, G9, G12 and G15 from the 3rd GAMA data release \citep{Liske2015}. These equatorial regions encompass roughly \SI{180}{\square\degree}, containing $\sim$\SI{180000} galaxies with sufficient quality redshifts. The magnitude-limited sample is essentially complete down to a magnitude of $r$ = 19.8. For our galaxy-galaxy lensing and clustering measurements, we use all GAMA galaxies in the three equatorial regions in the redshift range $0.15<z_{\rm l}<0.51$. As GAMA is essentially complete, the sample is equally weighted, such that $w_{\rm l}=1$ for all galaxies. We constructed random catalogues using the GAMA angular selection masks combined with an empirical smooth fit to the observed galaxy redshift distribution \citep{Blake2013}. We use the value for the RSD parameter from \citet{Blake2013} as $\beta=0.60 \pm 0.09$, which, we note encompasses a slightly different redshift range of $0.25< z_{\rm l} <0.5$, but still encompasses roughly 60\% of the galaxies in the sample, The use of this measurement is justified as $\beta$ varies slowly with redshift, therefore, any systematic uncertainty introduced by this choice is smaller than the statistical error of the measurement. This analysis measured $\beta$ similarly to the 2dFLenS case.

\subsection{Mocks}
We compute the full covariance between the different scales of the galaxy-galaxy lensing measurement using a large suite of $N$-body simulations, built from the Scinet Light Cone Simulations \citep[][SLICS]{HDvW2015} and tailored for weak lensing surveys. These consist of 600 independent dark matter only simulations, in each of which $1536^3$ particles are evolved within a cube of $505 h^{-1} \rm{Mpc}$ on a side and projected on 18 redshift mass planes between $0<z<3$. Light cones are propagated on these planes on $7745^3$ pixel grids and turned into shear maps via ray-tracing, with an opening angle of \SI{100}{\square\degree}. The cosmology is set to $\rm{WMAP9 + BAO + SN}$ \citep{Dunkley}, that is $\Omega_{\rm m}=0.2905$, $\Omega_{\Lambda}=0.7095$, $\Omega_{\rm b}=0.0473$, $h=0.6898$, $n_{\rm s}=0.969$ and $\sigma_8=0.826$.  These mocks are fully described by \citet{HDvW2015} and a previous version with a smaller opening angle of \SI{60}{\square\degree} was used in the KiDS analyses of \citet{Hildebrandt/etal:2017} and \citet{Joudaki2017}. 

Source galaxies are randomly inserted in the mocks, with a true redshift satisfying the KiDS DIR redshift distribution and a mock photometric redshift, $z_{\rm B}$. The source number density is defined to reflect the effective number density of the KiDS data. The gravitational shears are an interpolation of the simulated shear maps at the galaxy positions, while the distribution of intrinsic ellipticity matches a Gaussian with a width of  0.29 per component, closely matching the measured KiDS intrinsic ellipticity dispersion \citep{Amon2017, Hildebrandt/etal:2017}. 

To simulate a foreground galaxy sample, we populate the dark matter haloes extracted from the $N$-body simulations with galaxies, following a halo occupation distribution approach (HOD) that is tailored for each galaxy survey. The details of their construction and their ability to reproduce the clustering and lensing signals with the KiDS and spectroscopic foreground galaxy samples are described in \citep{JHD2018}. Here we summarise the strategy. Dark matter haloes are assigned a number of central and satellite galaxies based on their mass and on the HOD prescription. Centrals are placed at the halo centre and satellites are scattered around it following a spherically-symmetric NFW profile, with the number of satellites scaling with the mass of the halo. On average, about 9\% of all mock CMASS and LOWZ galaxies are satellites, a fraction that closely matches that from the BOSS data. The satellite fraction in the GAMA mocks is closer to 15\%. 

The CMASS and LOWZ HODs are inspired by the prescription of \citet{Alam2016} while the GAMA mocks follow the strategy of \citet{Smith2017}. In all three cases, we adjust the value of some of the best-fit parameters in order to enhance the agreement in clustering between mocks and data, while also closely matching the number density and the redshift distribution of the spectroscopic surveys. In contrast to the CMASS and LOWZ mocks, the GAMA mocks are constructed from a conditional luminosity function and galaxies are assigned an apparent magnitude such that we can reproduce the magnitude distribution of the GAMA data. For 2dFLenS, the LOWZ and CMASS mocks were subsampled to match the sparser 2dFLOZ and 2dFHIZ samples.

%%%%%%%%%%%%%%%%%%%%%%%%%%%%%%%%%%%%%%%%%%%%%%%%%%%%%%%

\section{Measurements} \label{sec:measurements} %\input measurements.tex
\subsection{Galaxy-galaxy annular surface density}
We compute the projected correlation function, $w_{\rm p}$ and the associated galaxy-galaxy annular surface density, $\Upsilon_{\rm gg}$, using the three-dimensional positional information for each of the five spectroscopic lens samples. We measure these statistics using random catalogues that contain $N_{\rm ran}$ galaxies, roughly 40 times the size of the galaxy sample, $N_{\rm gal}$, with the same angular and redshift selection. To account for this difference, we assign each random point a weight of $N_{\rm gal}/N_{\rm ran}$.

Adopting a fiducial flat $\Lambda$CDM WMAP cosmology \citep{Komatsu2011} with $\Omega_{\rm m}=0.27$, we estimate the 3D galaxy correlation function, $\xi_{\rm gg}(R,\Pi)$, as a function of comoving projected separation, $R$, and line-of-sight separation, $\Pi$, using the estimator proposed by \citet{LandySzlay1993},
\begin{equation}
\xi_{\rm gg}(R,\Pi)=\rm{\frac{dd-2dr+rr}{rr}}\, ,
\end{equation}
where $\rm{dd, rr}$ and $\rm{dr}$ denote the weighted number of pairs with a separation $(R,\Pi)$, where both objects are either in the galaxy catalogue, the random catalogue or one in each of the catalogues, respectively. 

\begin{figure*}
\centering
\includegraphics[width=\textwidth]{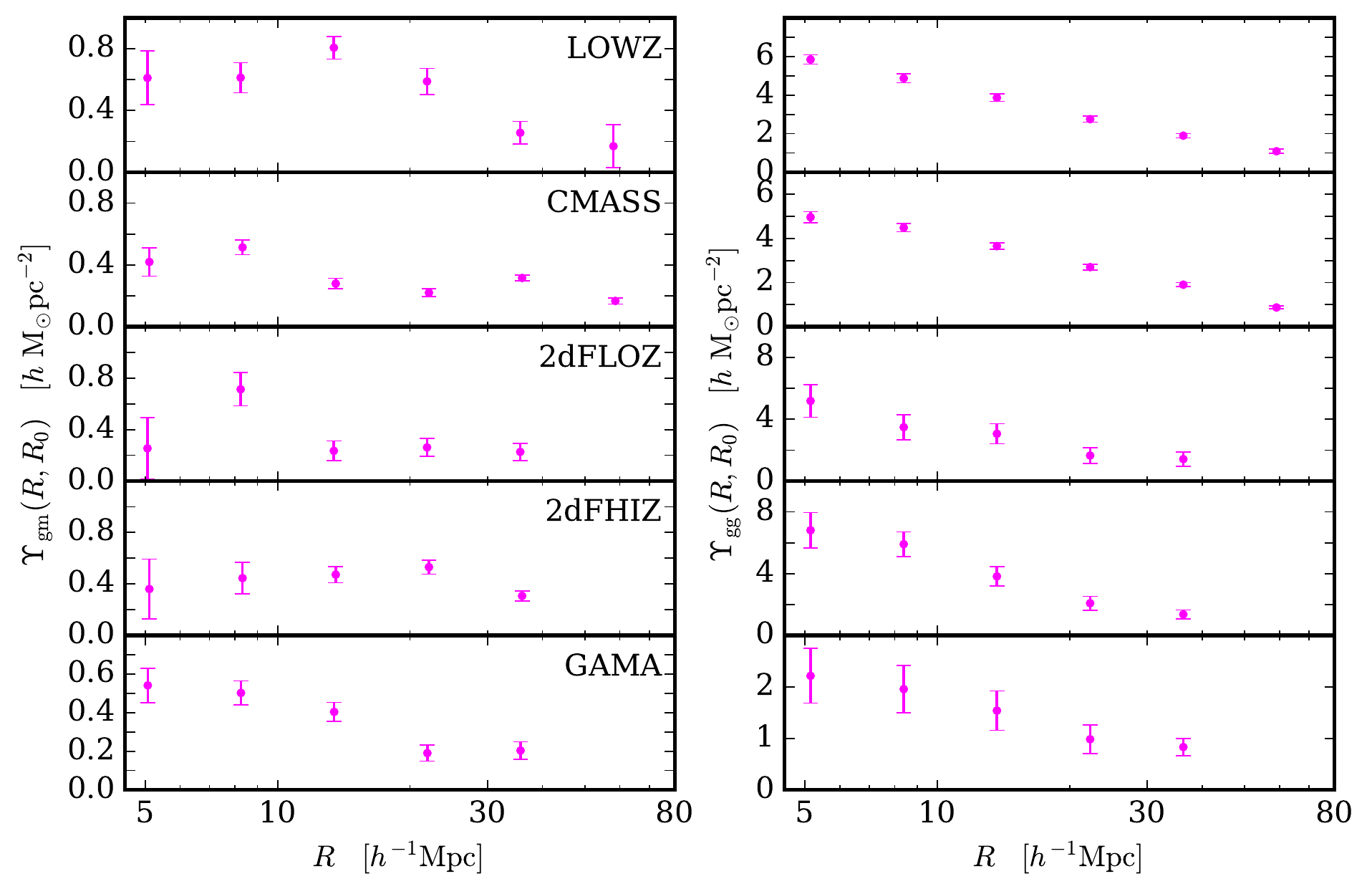}
\caption{ The galaxy-matter (left) and galaxy-galaxy (right) annular differential surface density measurements as a function of comoving scale, $\Upsilon_{\rm gm}(R,R_0=2.0\,  h^{-1} \rm{Mpc})$  and  $\Upsilon_{\rm gg}(R,R_0=2.0 h^{-1} \rm{Mpc})$, respectively, with LOWZ, CMASS, 2dFLOZ, 2dFHIZ and GAMA lens galaxy samples, from top to bottom. Scales below $R=5.0 \, h^{-1} \rm{Mpc}$ are not included in the analysis in order to remove regions where non-linear bias effects may enter, as well as to account for any bias introduced in the choice of $R_0$. For $\Upsilon_{\rm gm}$, errors are from simulations, while the error on $\Upsilon_{\rm gg}$  is determined from the propagation of a Jack-knife analysis.}
\label{fig:ups}
\end{figure*}

In order to obtain the projected correlation function, we combine the line-of-sight information by summing over 10 logarithmically-spaced bins in $\Pi$ from $\Pi=0.1$ to $\Pi=100 \, h^{-1}$Mpc, 
\begin{equation}
w_{\rm p}(R)=2 \sum_i \xi_{\rm gg}(R, \Pi_i) \Delta \Pi_i \, .
\end{equation}
We use 17 logarithmic bins in $R$ from $R=0.05$ to $R=100 \, h^{-1}$Mpc. The upper bound $\Pi_{\rm max}=100 \, h^{-1}$Mpc can potentially create a systematic error as $R$ approaches $\Pi_{\rm max}$ due to any lost signal in the range $\Pi>100 \, h^{-1}$Mpc, however the signal is negligible on these scales \am{and the measurement was robust to changes in this value for the level of precision of the analysis}. The error in $w_{\rm p}(R)$ is determined via a Jack-knife analysis, dividing the galaxy survey into 50 regions, ensuring a consistent shape and number of galaxies in each region. As such, the Jack-knife box size depends on the size of the survey at roughly \SI{1}{\square\degree} for GAMA and a few square degrees for the other lens samples.

We convert this measurement to a galaxy-galaxy annular differential surface density (ADSD), $\Upsilon_{\rm gg}$, following equation~\ref{eqn:upsgg}, where we define $R_0=2.0 h^{-1}$Mpc. A range of values of $R_0$ were tested between $1.0$ and $3.0 \, h^{-1}$Mpc and it was found that this choice affected only the first $R>R_0$ data point, but had no significant effect on the value of the $E_{\rm G}$ measurement over all other scales. As such, scales below  $R=5.0 h^{-1}$Mpc are not included. This choice removes regions where non-linear bias effects may enter, as well as account for any bias introduced by this choice of $R_0$. We determine $w_{\rm p}(R_0)$ via a power-law fit to the data in the range $R_0/3<R<3R_0$ and perform a linear interpolation to the measured $w_{\rm p}(R)$ in order to compute the integral in the first term. Any error in the interpolation for $w_{\rm p}(R_0)$ is ignored in the propagation of the Jack-knife error in $w_{\rm p}(R)$ to $\Upsilon_{\rm gg}$, as this contribution is only significant when $R \approx R_0$.

The right-hand panel of Figure~\ref{fig:ups} shows the measurements of $\Upsilon_{\rm gg}(R,R_0=2.0 h^{-1} \rm{Mpc})$ for each of the lens samples.

\subsection{Galaxy-matter annular surface density}
The galaxy-galaxy lensing estimator is defined as a function of angular separation in terms of the \textit{lens}fit weight of the sources, $w_{\rm s}$, the spectroscopic weight of the lenses, $w_{\rm l}$, and the tangential ellipticity of the source relative to the lens, $\epsilon_t$, as,
\be
\gamma_{\rm t}(\theta)=\frac{\sum^{\rm N_{pairs}}_{jk}  w_{\rm s}^j w_{\rm l}^k \epsilon_t^{jk}}{\sum^{\rm N_{pairs}}_{jk} w_{\rm s}^j w_{\rm l}^k} \, .
\ee
This statistic is measured with a selection function such that only source-lens galaxy pairs within a separation in the interval $[\theta,\theta+\Delta\theta]$ are probed. For this measurement we employ the \textsc{treecorr} software of \citet{Jarvis2004}, but we have performed consistency checks using the \textsc{athena} software of \citet{Kilbinger2014}.

Lens galaxies were selected by their spectroscopic redshift, $z_{\rm l}$, into $N_{\rm z}$ redshift `slices' of width $\Delta z_{\rm l}=0.01$ between $0.15<z_{\rm l}<0.43$ for LOWZ and 2dFLOZ,  $0.15<z_{\rm l}<0.51$ for GAMA and  $0.43<z_{\rm l}<0.7$ for CMASS and 2dFHIZ. For each slice of the lens catalogue, the tangential shear was measured in 17 logarithmic angular bins where the minimum and maximum angles were determined by the redshift of the lens slice as $\theta= R/\chi(z_{\rm l})$ in order for all slice measurements to satisfy minimum and maximum comoving projected radii from the lens of $R=0.05$ and $R=100 \, h^{-1}$Mpc. For each slice measurement, the source sample is limited to those behind each lens slice, in order to minimise the dilution of the lensing signal due to sources associated with the lens. The selection is made using the  $z_{\rm B}$ photometric redshift estimate as $z_{\rm B}>z_{\rm l}+0.1$, which was deemed most optimal in Appendix D of \citet{Amon2017}. The redshift distribution for each source subsample,  $N(z_{\rm s})$, is computed with the DIR method for each spectroscopic slice.

The inverse comoving critical surface mass density is calculated per source-lens slice following equation~\ref{eqn:scviola} and~\ref{eqn:scconvert} as,
\begin{equation}
\label{eqn:sigcrit}
\centering
\overline{\Sigma}_{\rm c, com}^{-1}[z_{\rm l},{N}(z_{\rm s})] = \frac{4\pi G (1+z_{\rm l}) \chi(z_{\rm l})}{c^2} \int^{\infty}_{z_{\rm l}} dz_{\rm s} \, {N}(z_{\rm s}) \Big[1- \frac{\chi(z_{\rm l})}{\chi(z_{\rm s})}\Big] \, ,
\end{equation}
where  $\overline{\Sigma}_{\rm c, com}^{-1}[z_{\rm l},{N}(z_{\rm s})]$ is the inverse critical surface mass density at $z_{\rm l}$, averaged over the entire source redshift distribution, ${N}(z_{\rm s})$, normalised such that $\int {N}(z_{\rm s}) {\rm d}z_{\rm s}=1$. $\chi(z_{\rm l})$ and $\chi(z_{\rm s})$ are the comoving distances to the lens and source galaxies, respectively.
Again, we adopt a fiducial flat $\Lambda$CDM WMAP cosmology with $\Omega_{\rm m}=0.27$. Our motivation for this choice is to ensure an unbiased measurement by choosing a cosmology with a value for the matter density which lies between the values favoured by KiDS and Planck. This also ensures consistency with the fiducial cosmological model adopted for the RSD analyses of the BOSS and GAMA analysis, which would be subject to Alcock-Paczynski distortion in different models \citep{AP1979}.  Adopting the different value of $\Omega_{\rm m}$ preferred by the Planck and KiDS analyses would not produce a significant change in our measurements compared to their statistical errors. 

The estimator for the excess surface mass density is defined as a function of the projected radius and the spectroscopic redshift of the lens as a combination of the inverse critical surface mass density and the tangential shear,
\be
\Delta\Sigma_{\rm com}(R,z_{\rm l})=\frac{\gamma_{\rm t}[\theta=R/\chi(z_{\rm l})]}{\overline{\Sigma_{\rm c,com}^{-1}}[z_{\rm l},N(z_{\rm s})]} \, .
\ee

We calculate the tangential shear and the differential surface mass density, $\Delta\Sigma (R)$, for each of the $N_{\rm z}$ lens slices and stack these signals to obtain an average differential surface mass density, weighted by the number of pairs in each slice as,
\be
\centering
\overline{\Delta\Sigma}_{\rm com}(R)=\frac{\sum^{N_{\rm z}}_i [\gamma_{\rm t}(R/\chi_{\rm l})/\overline{\Sigma_{\rm c,com}^{-1}}]^i n_{\rm{pairs}}^i \ K^i}{\sum^{N_{\rm z}}_i n_{\rm{pairs}}^i} \, ,
\ee
where we include a shear calibration for each redshift slice $K^i$, where,
\begin{equation}
\centering
K^i=\frac{\sum_{\rm s} w_{\rm s} (1+m_{\rm s})}{\sum_{\rm s} w_{\rm s}} \, ,
\end{equation}
and $m_{\rm s}$ is the multiplicative bias per source galaxy as derived in \citet{fenech-conti/etal:2016}.

While it is common to apply a `boost factor' in order to account for source galaxies that are physically associated with the lenses that may bias the tangential shear measurement, we show in \citet{Amon2017} that this signal is negligible for our lens samples and redshift selections for scales beyond $R=2.0 \, h^{-1}$Mpc. As we only probe larger scales than this, we do not apply this correction.
The excess surface mass density was also computed around random points in the areal overlap. This signal has an expectation value of zero in the absence of systematics. As demonstrated by \cite{Singh2016}, it is important that a random signal, $\Delta\Sigma_{\rm{rand}}(R)$, is subtracted from the measurement in order to account for any small but non-negligible coherent additive bias of the galaxy shapes and to decrease large-scale sampling variance. The random signals were found to be consistent with zero for each lens sample \citep{Amon2017}. 

The error in the measurements of $\overline{\Delta\Sigma}(R)$ combines in quadrature the uncertainty in the random signal and the full covariance determined from simulations, as described in Section~\ref{sec:cov}. A bootstrap analysis of the redshift distribution in \citet{Hildebrandt/etal:2017} revealed that this uncertainty is negligible compared to the lensing error budget for our analysis, as was also found in \citet{Dvornik2017}.

We convert the measurements of the excess surface mass density and its covariance into the galaxy-matter ADSD, $\Upsilon_{\rm gm}$, following equation~\ref{eqn:upsgm}, with $R_0=2.0 h^{-1}$Mpc. Similarly to the case of $\Upsilon_{\rm gg}$, we determine $\Delta\Sigma(R_0)$ by a power-law fit to the data and ignore any error on this interpolation.

The left-hand panel of Figure~\ref{fig:ups} shows the measurements of $\Upsilon_{\rm gm}(R,R_0=2.0 h^{-1} \rm{Mpc})$ for the cross-correlation with each of the lens samples. The ranges plotted, that is $5<R<60 h^{-1}$ for LOWZ and CMASS and $5<R<40 h^{-1}$Mpc for 2dFLOZ, 2dFHIZ and GAMA, represent the scales where the assumption of linear bias holds and where we trust the Jack-knife error analysis for the clustering measurements in the cases of 2dFLenS and BOSS. \am{We conservatively choose the minimum fitted scale to be roughly double $R_0$. Our measurements of $E_{\rm G}$ are insensitive to this choice.} These are the scales used in the measurements and fits of $E_{\rm G}(R)$. We note that the shapes and amplitudes of the lensing profiles on the left-hand side of Figure~\ref{fig:ups} differ as the lens galaxy samples vary in flux limits, redshift and for the case of comparison between BOSS and 2dFLenS, completeness.

\subsection{Covariance for $E_{\rm G}$} \label{sec:cov}
We measure the galaxy-galaxy lensing signal using the ensemble of $N_{\rm sim} = 600$ N-body simulations of source and lens catalogues with the same pipeline applied to the data. We construct the covariance matrix from these measurements by scaling the resulting covariance by  $100 \rm{deg}^2/A_{\rm eff}$ for each region, where $A_{\rm eff}$ represents the effective overlap area of the surveys, as listed in Table~\ref{tab:spec} \citep{Schneider2002}.  The covariance, $C$, between the measurements at transverse scales $R^i$ and $R^j$, is computed as,
\begin{equation}
\begin{split}
\hat{C}^{i,j}=\hat{C}[\Upsilon_{\rm gm}(R^i),\Upsilon_{\rm gm}(R^j)]=\frac{1}{N_{\rm sim} -1} \times \\
\Big[ \sum_{k=1}^{N_{\rm{sim}}} \big(\Upsilon^k_{\rm gm}(R^{i}) -  \overline{\Upsilon_{\rm gm}}(R^i) \big) \big(\Upsilon^k_{\rm gm}(R^{j}) & -\overline{\Upsilon_{\rm gm}}(R^j) \big) \Big] \, ,
\label{eqn:egcov}
\end{split}
\end{equation}
where $\Upsilon_{\rm gm}^k(R^i)$ is measured for the $k$th mock catalogue and $\overline{\Upsilon_{\rm gm}}(R^i)$ is the average over all mocks.

Under the assumption that the three measurements that we combine to estimate $E_{\rm G}$ are independent, we estimate the covariance matrix as a combination of the covariance of the galaxy-galaxy lensing measurement estimated from N-body simulations, the Jack-knife covariance for the clustering measurement and the error in the $\beta$ parameter, which modifies all scales and therefore folds through as a scalar of amplitude $\sigma_{\beta}$ multiplied by a unit matrix. Using the chain rule for ratios, we obtain:
\begin{equation}
\frac{\hat{C}(E_{\rm G})^{i,j}}{E_{\rm G}^iE_{\rm G}^j}=\frac{\hat{C}(\Upsilon_{\rm gm})^{i,j}}{\Upsilon_{\rm gm}^i\Upsilon_{\rm gm}^j} 
+  \frac{\hat{C}(\Upsilon_{\rm gg})^{i,j}}{\Upsilon_{\rm gg}^i\Upsilon_{\rm gg}^j} + \Big( \frac{\sigma_{\beta}}{\beta} \Big)^2  \, . 
\label{eqn:EGcov}
\end{equation}

In Appendix~\ref{app} we show the covariance matrix for each of the additive components of equation~\ref{eqn:EGcov} and thereby demonstrate that the error in the clustering measurement is subdominant compared to the galaxy-galaxy lensing measurement, justifying our use of a Jack-knife approach rather than mock analysis for this clustering component. For the cases of BOSS and 2dFLenS analyses, the lensing measurements use a small fraction of the total area used for the clustering measurement. This justifies our choice to neglect the cross-covariance between the two measurements and assume that the lensing, clustering and RSD measurements are independent. In Appendix~\ref{app} we discuss the case of GAMA and the appropriateness of these assumptions, given that the lensing area is not significantly smaller than the clustering area. The errors in the measurements of the RSD parameter, $\beta$, are drawn from the literature and quoted in Table~\ref{tab:spec}. 

The inverse of this covariance matrix is used in the model fitting of $E_{\rm G}(R)$. Whilst we consider our measurement of $\hat{C}(E_{\rm G})$ from the simulations to be an unbiased estimator of the true covariance matrix $\hat{C}$, it will have an associated measurement noise as it is constructed from a finite number of semi-independent realisations. As such, $\hat{C}^{-1}$ is not an unbiased estimate of the true inverse covariance matrix. We correct for this bias due to its maximum-likelihood estimation \citep{Hartlap2007} as $C^{-1}=\alpha \ \hat{C}^{-1}$, where
\begin{equation}
\alpha =\frac{N_{\rm sim} - N_{\rm bin} - 2}{N_{\rm sim}-1}\, ,
\end{equation}
and $N_{\rm bin}$ is the number of data bins used in the fit. This correction is valid under the condition that the number of simulations exceeds the number of data bins with $N_{\rm bin}/N_{\rm sim} < 0.8$. In this case of a large number of simulations, the correction by \citet{Hartlap2007} gives the same results as the more robust correction of \citet{Sellentin2016}.

\begin{figure*}
\centering
\includegraphics[width=\textwidth]{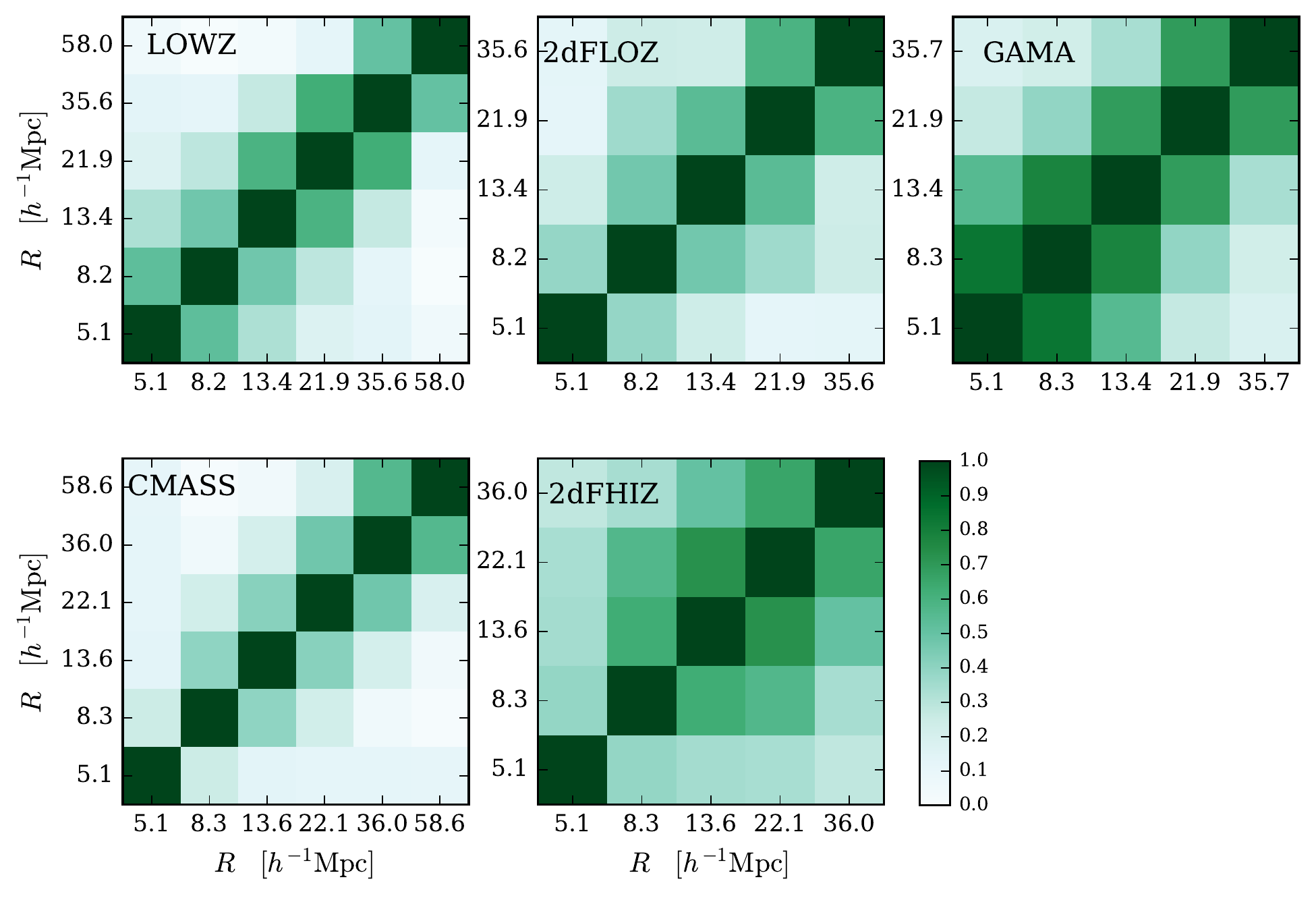}
\caption{\label{fig:covs} Correlation coefficients, $\zeta$, defined by equation~\ref{eqn:cor}, of the covariance matrix of the $E_{\rm G}$ measurements, determined with each of the five lens samples. These are computed as a combination of the $\Upsilon_{\rm gm}$ covariance determined from the scatter across the 600 simulation line-of-sights, the Jack-knife covariance of $\Upsilon_{\rm gg}$ and the uncertainty on the RSD parameter, $\beta$.}
\end{figure*}

The correlation matrix for $E_{\rm G}$ is determined from the covariance as,
\be
\label{eqn:cor}
\zeta(E_{\rm G})^{i,j}=\frac{\hat{C}(E_{\rm G})^{ij}}{\sqrt{\hat{C}(E_{\rm G})^{ii}\hat{C}(E_{\rm G})^{jj}}} \, .
\ee
Figure~\ref{fig:covs} illustrates the correlation matrices of the measurements with each of the five lens samples. The correlation between different physical scales is most significant for cross-correlations with GAMA and 2dFHIZ and is non-negligible for the high-redshift samples.

%%%%%%%%%%%%%%%%%%%%%%%%%%%%%%%%%%%%%%%%%%%%%%%%%%%%%%%

\section{Cosmological Results} \label{sec:results} %\input results.tex

\begin{figure*}
\centering
\includegraphics[width=\textwidth]{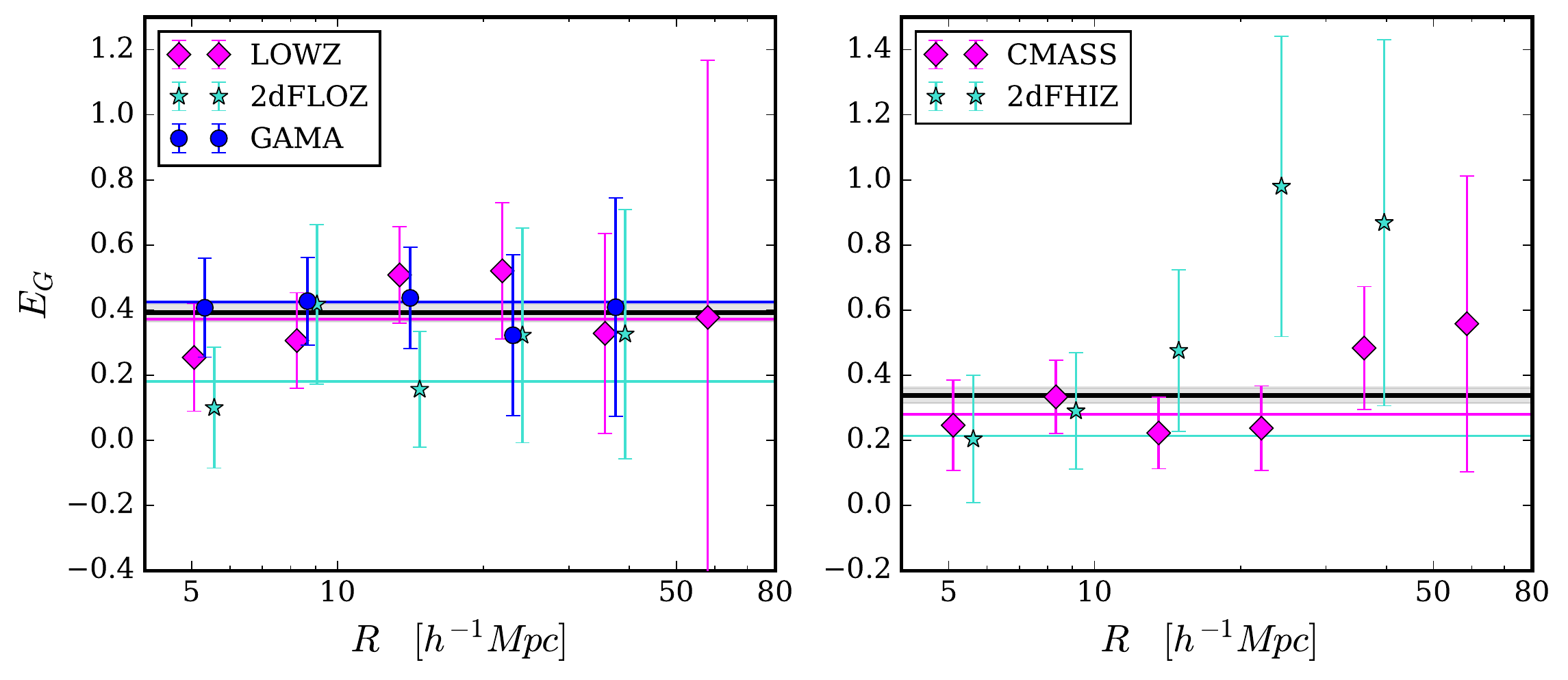}
\caption{The $E_{\rm G}$ statistic, $E_{\rm G}(R)$, computed using KiDS-450 data combined with low-redshift spectroscopic lenses from GAMA (blue) in the range $0.15<z_{\rm l}<0.51$ and from 2dFLOZ (turquoise) and LOWZ (pink) in the range $0.15<z_{\rm l}<0.43$ in the left-hand panel and high-redshift lenses spanning  $0.43<z_{\rm l}<0.7$ from CMASS (pink) and 2dFHIZ (turquoise) in the right-hand panel. Data points are offset on the $R$-axis for clarity. The solid black line denotes the GR prediction for a KiDS+2dFLenS+BOSS cosmology with $\Omega_{\rm m}=0.243 \pm 0.038$. The coloured lines denote the best-fit scale-independent models to the measurements.}
\label{fig:eg}
\end{figure*}

We combine the lensing and clustering measurements with the redshift-space distortion parameters following equation~\ref{eqn:eg}.
We note that while our analysis includes the uncertainty related to each redshift-space distortion measurement, any potential remaining systematic errors on $\beta$ could bias the $E_{\rm G}$ result. 
Figure~\ref{fig:eg} shows our measurements of $E_{\rm G}(R)$ for the low-redshift lens samples (left) and the high-redshift lens samples (right). The black-line represents the GR prediction, determined with the KiDS+2dFLenS+BOSS cosmology measured by \citet{Joudaki2017}, that is, with a matter density today of $\Omega_{\rm m}(z=0)=0.243 \pm 0.038$. The coloured lines denote the best-fit scale-independent model, as determined by the minimum chi-squared using the covariance defined in equation~\ref{eqn:egcov}. 

The mean and 1$\sigma$ error in the scale-independent best fit to the measurements, as shown in Figure~\ref{fig:eg}, are quoted for each lens sample in Table~\ref{tab:eg}. The $\chi^2_{\rm min}$ for each of the analyses are quoted in the Table. We note that the $\chi^2_{\rm min}$ for the analysis with GAMA is slightly lower than expected for 4 degrees of freedom. In Appendix~\ref{app} we investigate the effect of the covariance on these fits for each of the lens samples. We argue that for the analysis with GAMA, the clustering error is overestimated due to the size of the Jackknife region and causes an overestimation of the uncertainty of $E_{\rm G}$, but is unlikely to bias the fit. 

\begin{figure}
\centering
\includegraphics[width=\columnwidth]{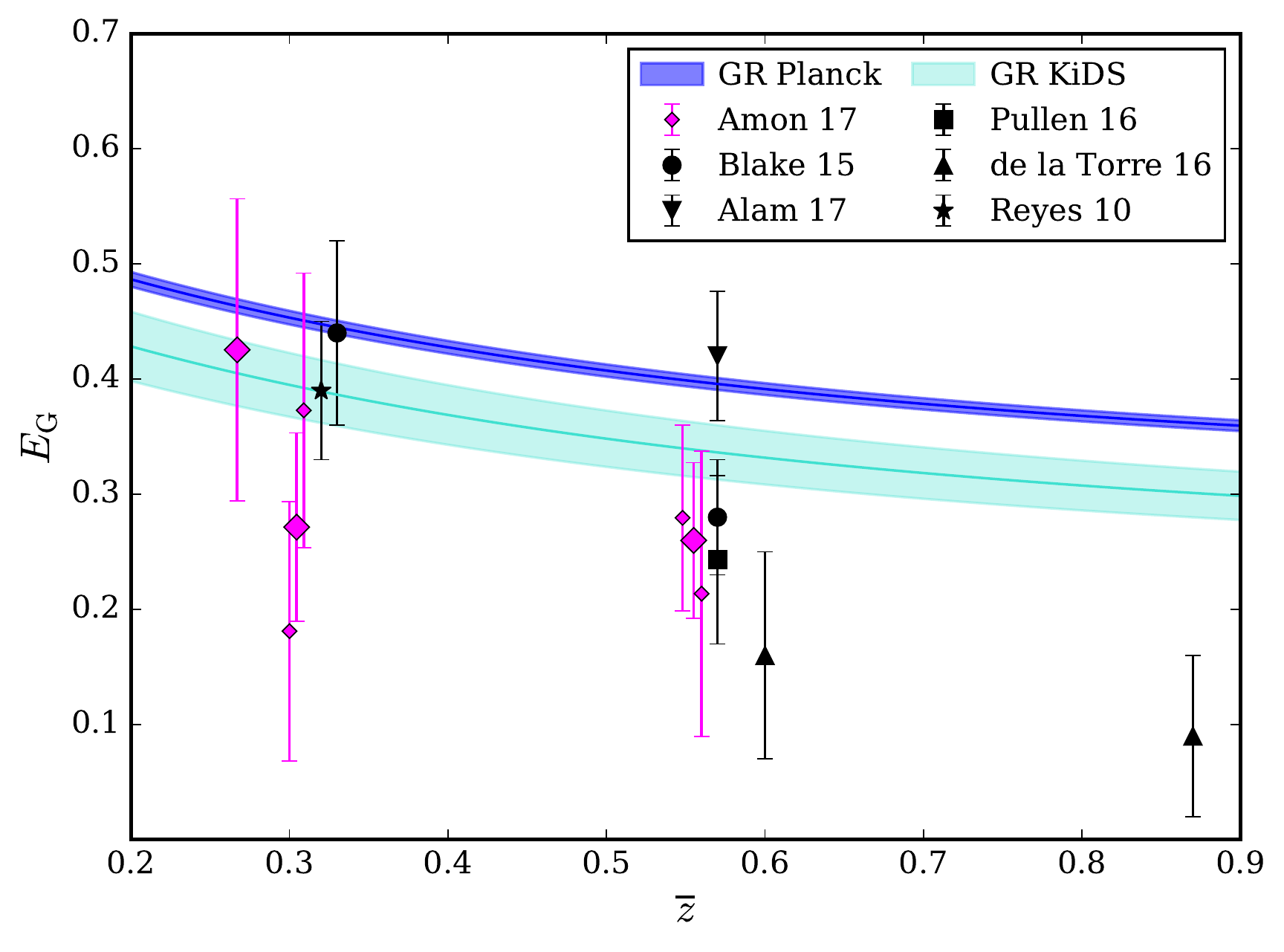}
\caption{The scale-independent fit to the $E_{\rm G}(R)$ measurements shown in Figure~\ref{fig:eg}, now plotted as a function of the mean redshift of the spectroscopic lens sample, $E_{\rm G}(\overline{z})$. From left to right, the smaller pink data points represent the fits to the measurements computed using KiDS-450 combined with 2dFLOZ, LOWZ, CMASS and 2dFHIZ. The errorbars denote the 1$\sigma$ uncertainty on the fit to the data. The larger pink data points represent the fit to the measurement with GAMA, as well as the combination of the independent fits from  2dFLOZ+LOWZ and 2dFHIZ+CMASS. The blue region denotes the 68\% confidence region of GR for a Planck (2016) cosmology while the turquoise region represents that for the KiDS+2dFLenS+BOSS cosmology. }
\label{fig:egvsz}
\end{figure}

In Figure~\ref{fig:egvsz} we plot the fits to our measurements as a function of the mean redshift of the spectroscopic sample in pink. BOSS and 2dFLenS are in different parts of the sky and therefore give independent measurements, which we find to be consistent with each other at roughly 1.5$\sigma$. As such, we combine the measurements at the same redshift using inverse-variance weighting and find $E_{\rm G}(\overline{z}=0.305)=0.27 \pm 0.08$ for the combination of LOWZ+2dFLOZ and $E_{\rm G}(\overline{z}=0.554)=0.26 \pm 0.07$ for the combination of CMASS+2dFHIZ. These combinations are denoted by larger pink data points. Alongside the results of this analysis, we plot existing measurements of $E_{\rm G}$ in black \citep{Reyes2010,Blake2016eg,Pullen2016,Alam2016,sdlt2016}. In light of the current tension between CMB temperature measurements from Planck and KiDS lensing data, we plot two GR predictions using both the preferred Planck cosmology \citep{Planck2016} and the KiDS+2dFLenS+BOSS cosmology \citep{Joudaki2017}. The Planck cosmology is drawn from \citet{Planck2016}, with $\Omega_{\rm m}(z=0)=0.308 \pm 0.009$. The 68\% confidence regions are denoted by the shaded regions. 

While the \citet{Reyes2010} result and the low-redshift \citet{Blake2016eg} measurement of $E_{\rm G}$ are consistent with both the GR predictions, the high-redshift measurements show variation. \citet{Alam2016} found their high-redshift measurement of the $E_{\rm G}$ statistic to be consistent with both cosmologies. On the other hand, \citet{sdlt2016}, the high-redshift measurement from \citet{Blake2016eg} and the CMB-lensing \cite{Pullen2016} measurement find values of the statistic that are more than 2$\sigma$ low when compared to the Planck GR prediction. Notably, the highest-redshift $E_{\rm G}$ measurements by \citet{sdlt2016} are in tension with a KiDS+2dFLenS+BOSS GR prediction. The $E_{\rm G}$ statistic was motivated solely as a test of GR, but a choice of cosmology has to be made in computing this prediction. As Figure~\ref{fig:egvsz} shows, this choice has a significant impact on conclusions. Interestingly, in general, our $E_{\rm G}$ measurements and previous measurements from the literature prefer lower values of the matter density parameter such as those constrained by KiDS+2dFLenS+BOSS.  

\begin{table}
\caption{\label{tab:eg} The scale-independent fit to the $E_{\rm G}(R)$ measurements and the 1$\sigma$ error on the parameter in the fit, along with the minimum $\chi^2$ value and number of degrees of freedom (d.o.f.), for the analyses using each of the spectroscopic samples.} 
\begin{center}
\begin{tabular}{lcccc}
 \hline
Spec. sample & $E_{\rm G}$ & $\chi^2_{\rm min}$ & d.o.f.\\
 \hline
LOWZ & 0.37$\pm$0.12 & 2.8 & 5 \\
2dFLOZ& 0.18$\pm$0.11 & 2.1 & 4 \\ 
CMASS &  0.28$\pm$0.08 & 3.2 & 5 \\
2dFHIZ &  0.21$\pm$0.12 & 3.4 & 4 \\ %\cline{2-5}
GAMA &  0.43$\pm$0.13 & 0.8 & 4 \\
\hline
\end{tabular}
\end{center}
\end{table}

\begin{figure}
\centering
\includegraphics[width=\columnwidth]{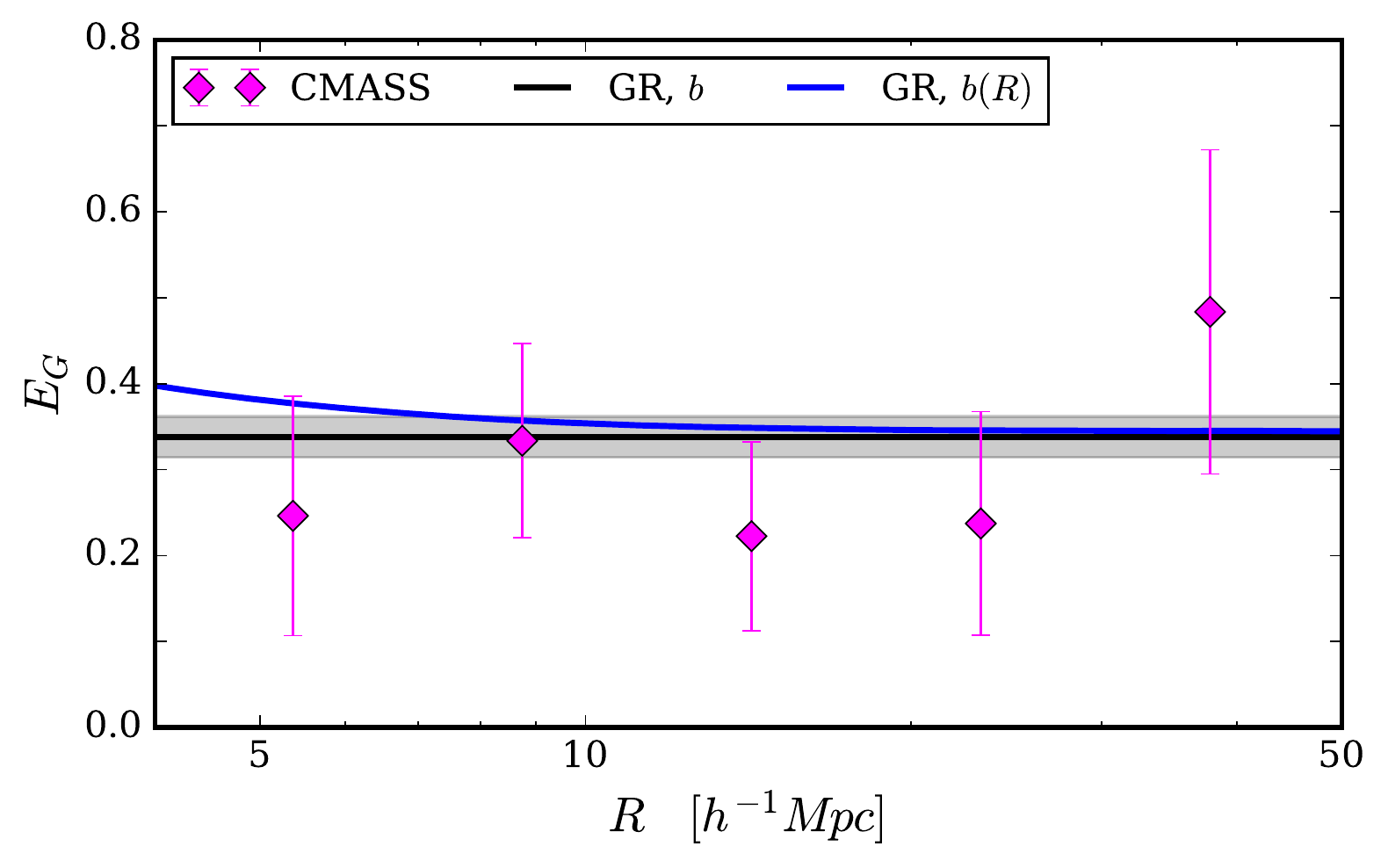}
\caption{The effect of a scale-dependent galaxy bias on the predictions of the $E_{\rm G}$ statistic. We show $E_{\rm G}(R)$, computed with CMASS spectroscopic lenses (pink) plotted with the GR prediction for a KiDS+2dFLenS+BOSS cosmology with the fiducial scale-independent bias model, $b$, (black) and a scale-dependent bias model, $b(R)$ (blue).}
\label{fig:bias}
\end{figure}

%In Figure~\ref{fig:bias}, we investigate our assumption of scale-independent bias. We show the prediction for $E_{\rm G}(R)$ in GR and with a KiDS cosmology, assuming the scale-dependent galaxy bias model given by the 2dF Galaxy Redshift Survey (2dFGRS) in \cite{Cole2005}. Alongside, we plot the measurement with GAMA galaxies, as this sample is most similar in redshift to 2dFGRS. The effect of including this scale-dependence is shown to be minimal in comparison with the errors on our measurements, which provides support for our assumption of linear bias on the projected scales in question. We do however caution that the bias model of \cite{Cole2005} is fit to a somewhat less bright galaxy population than the BOSS LRG samples, and this prediction therefore only serves to illustrate the expected low-level impact of scale-dependent bias on our analysis.
In Figure~\ref{fig:bias}, we investigate our assumption of scale-independent bias. We show the prediction for $E_{\rm G}(R)$ in GR and with a KiDS cosmology, assuming a scale-dependent galaxy bias model using CMASS Halo Occupation Distribution parameters from \citet{More2015}. Alongside, we plot the measurement with CMASS galaxies. The effect of including this scale-dependence is shown to be minimal in comparison with the errors on our measurements, which provides support for our assumption of linear bias on the projected scales in question. We do however caution that the bias model is fit to a marginally fainter galaxy population than the 2dFLenS LRG samples, and this prediction therefore only serves to illustrate the expected low-level impact of scale-dependent bias on our analysis. As GAMA contains less bright galaxies than CMASS, we assume that the effect of scale dependent bias is smaller for this case. The value of $E_G$ in GR with scale-dependent bias deviates from the scale-independent prediction by at most 10 percent over the scales in which we are interested.

\begin{figure}
\centering
\includegraphics[width=\columnwidth]{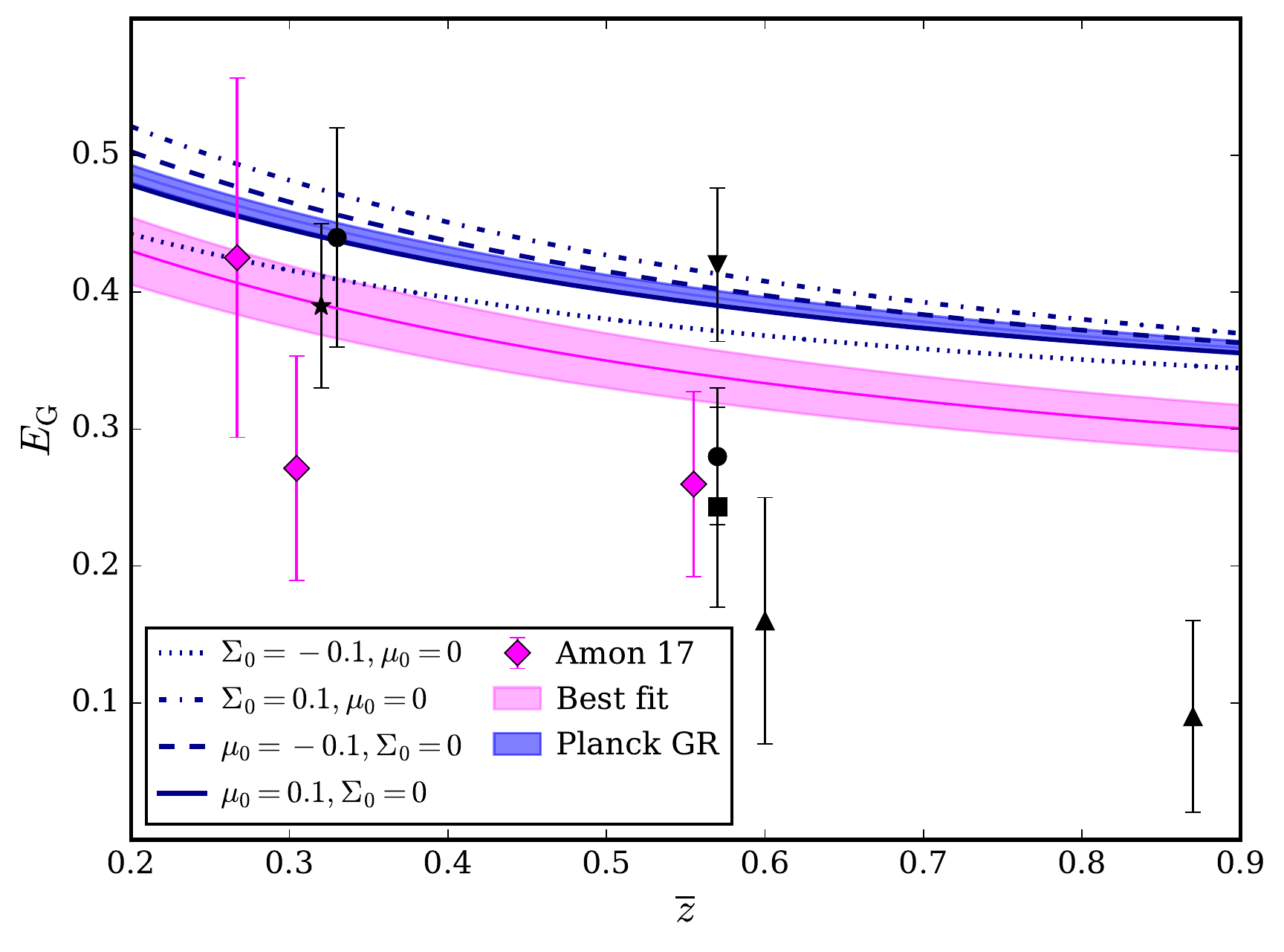}
\caption{Fits to the measurements of the $E_{\rm G}$ statistic, $E_{\rm G}(\overline{z})$ measured with KiDS combined with GAMA, LOWZ+2dFLOZ and CMASS+2dFHIZ data compared to the theoretical predictions of the statistic with different gravity models for Planck (2016) cosmology. The blue shaded region represents the prediction from GR, while the lines denote the theoretical predictions for modifications to gravity in a $(\Sigma_0,\mu_0)$ parametrisation with different departures from $(0,0)$. The pink shaded region shows the best-fit model for our $E_{\rm G}$ measurements with that from \citet{Reyes2010}, the low-redshift \citet{Blake2016eg} and \citet{Alam2016}.}
\label{fig:eg_KMG}
\end{figure}

Figure~\ref{fig:eg_KMG} compares our three measurements to predictions of $E_{\rm G}(z)$ with modifications to GR in the phenomenological $\{\mu_0,\Sigma_0 \}$ parametrisation described in Section~\ref{sec:MG}, with a Planck cosmology. We show variations to either $\mu_0$ or $\Sigma_0$ and find that $E_{\rm G}$ is more sensitive to the latter. 

\am{Using a combination of the measurements from this analysis and those from the literature, we calculate the best-fit prediction for $E_{\rm G}$ and therefore for the matter density parameter today. We perform this fit under the assumption of GR and compute the matter density parameter following equation~\ref{EG_theory}, $E_{\rm G}= \Omega_{\rm m}(z=0)/f(z)$, and using the relation $f(z)=\Omega_{\rm m}(z)^{0.55}$. We exclude the data points from \citet{sdlt2016} from our fit on the basis that they conclude that their measurement of $E_{\rm G}$ is biased due to systematics. Furthermore, we do not include the high-redshift result of \citet{Blake2016eg} in our fit as \citet{Alam2016} is a more recent analysis with the same datasets. We choose to only combine measurements that use galaxy-galaxy lensing, instead of CMB lensing and so we do not count the result of \citet{Pullen2016}. We find $\Omega_{\rm m}(z=0)=0.25\pm0.03$ with a $\chi^2_{\rm min}=6.3$ for 5 degrees of freedom. A model with this value for the matter density parameter is represented as the pink shaded region in Figure~\ref{fig:eg_KMG}.}

%%%%%%%%%%%%%%%%%%%%%%%%%%%%%%%%%%%%%%%%%%%%%%%%%%%%%%%

\section{Summary and Outlook}  \label{sec:conc}  %\input conc.tex
We have performed a new measurement of the $E_{\rm G}$ statistic. This was achieved by using measurements of redshift-space distortions in 2dFLenS, GAMA and BOSS galaxy samples and combining them with measurements of their galaxy-galaxy lensing signal, made using the first \SI{450}{\square\degree} of the Kilo-Degree Survey. Our results are consistent with the prediction from GR for a perturbed FRW metric, in a $\Lambda$CDM Universe with a KiDS+2dFLenS+BOSS cosmology, given by \citet{Joudaki2017}.

In particular, we determine $E_{\rm G}(z=0.267)= 0.43\pm0.13 $ using GAMA and averaging over scales $5<R<40 h^{-1}$Mpc, $E_{\rm G}(z=0.305)= 0.27\pm0.08$ using a combination of LOWZ and 2dFLOZ and averaging over scales $5<R<60 h^{-1}$Mpc and $E_{\rm G}(z=0.554)= 0.26\pm0.07$ using a combination of CMASS and 2dFHIZ over scales $5<R<60 h^{-1}$Mpc.  To obtain these constraints we fit a constant $E_{\rm G}$ model and incorporate the covariance matrix determined by a combination of the lensing covariance measured using a suite of N-body simulations, the clustering covariance determined from a Jack-knife analysis and the uncertainty on the RSD parameter, while neglecting the \am{cross-covariances} between the clustering and lensing measurements. In order to down-weight small scales where systematic corrections become significant and baryonic physics might have an effect, we suppress small-scale information from $R<R_0=2.0 h^{-1}$Mpc using annular statistics for the projected clustering and differential surface mass density and find that above $R=5.0 h^{-1}$Mpc, our results are insensitive to the choice of $R_0$, consistent with previous analyses. 

We show that while $E_{\rm G}$ is traditionally regarded a test of GR gravity, the robustness of this test is hindered by the uncertainty in the background cosmology, as illustrated by the current tensions between cosmological parameters defined by CMB temperature measurements from Planck and state of the art lensing data. While previous measurements of $E_{\rm G}$ \citep{Reyes2010, Blake2016eg,Pullen2016,sdlt2016} have reported low measurements when compared to a GR prediction with a Planck cosmology \citep{Planck2016}, similar to our findings, these apparent deviations are mostly resolved by a lower $\Omega_{\rm m}$ cosmology. Using our measurements combined with literature measurements from \citet{Reyes2010, Blake2016eg} and \citet{Alam2016}, we find that the best fit model for $E_{\rm G}$ uses a cosmology with a matter density as $\Omega_{\rm m}(z=0)=0.25\pm0.03$ with a $\chi^2_{\rm min}=6.3$ for 5 degrees of freedom. 
We present calculations of $E_{\rm G}$ in a 2-parameter modified gravity scenario and show that 10\% changes in the metric potential amplitudes produce smaller differences in the predicted $E_{\rm G}$ than changing $\Omega_{\rm m}$ between the values favoured by Planck and KiDS.

With Hyper Supreme-Cam \citep{HSC2017}, as well as the advent of next-generation surveys like LSST\footnote{\url{http://www.lsst.org/}}, Euclid\footnote{\url{http://sci.esa.int/euclid/}}, WFIRST\footnote{\url{http://wfirst.gsfc.nasa.gov/}}, 4MOST\footnote{\url{https://www.4most.eu/cms/}} and DESI\footnote{\url{http://desi.lbl.gov/}} 
surveys, these cross-correlations and joint analyses will become increasingly important in testing our theories of gravity \citep{Rhodes2013}. However, we caution that measurements of the $E_{\rm G}$ statistic cannot be conducted as consistency checks of GR until the tension in cosmological parameters is resolved. 

%%%%%%%%%%%%%%%%%%%%%%%%%%%%%%%%%%%%%%%%%%%%%%%%%%%%%%%

\section{\texorpdfstring{2\MakeLowercase{d}FL\MakeLowercase{en}S}{2dFLenS} Data Release}  \label{sec:2df}  
Simultaneously with this paper, full data catalogues from 2dFLenS (a 
subset of which are used in our current analysis) will be released via 
the website {\tt http://2dflens.swin.edu.au}.  The construction of these 
catalogues is fully described by \citet{Blake/etal:2016}, and we briefly 
summarize the contents of the data release in this section.

\begin{itemize}

\item The {\bf final 2dFLenS redshift catalogue} contains the $70{,}079$ 
good-quality spectroscopic redshifts obtained by 2dFLenS across all 
target types.  These include $40{,}531$ Luminous Red Galaxies (LRGs) 
spanning redshift range $z < 0.9$, $28{,}269$ redshifts which form a 
magnitude-limited nearly-complete galaxy sub-sample in the $r$-band 
magnitude range $17 < r < 19.5$, and a number of other target classes 
including a point-source photometric-redshift training set, compact 
early-type galaxies, brightest cluster galaxies and strong lenses.

\item The selection function of the Luminous Red Galaxy sub-samples has 
been determined, as described by \citet{Blake/etal:2016}.  The data release 
contains {\bf LRG data and random catalogues} for low-redshift ($0.15 < 
z < 0.43$) and high-redshift ($0.43 < z < 0.7$) LRGs in the KiDS-South 
and KiDS-North regions, after merging the different LRG target 
populations.

\item {\bf Mock data and random catalogues} for 2dFLenS LRGs were 
constructed by applying a Halo Occupation Distribution to an N-body 
simulation, as described by \citet{Blake/etal:2016}. The data release 
contains 65 mocks sub-sampled with the 2dFLenS selection function; the 
mock random catalogues slightly differ from the data random catalogues 
owing to approximations in mock generation (that are unimportant for 
cosmological applications).

\end{itemize}

%%%%%%%%%%%%%%%%%%%%%%%%%%%%%%%%%%%%%%%%%%%%%%%%%%%%%%%

\section*{Acknowledgements}
We thank the anonymous referee for careful and thorough comments, Shadab Alam for his advice and for kindly sharing measurements of the RSD parameter from BOSS DR12 and Arun Kannawadi Jayaraman and Vasiliy Demchenko for useful comments. Furthermore, we thank the referee for careful and thorough comments.
AA, CH, MA and SJ acknowledge support from the European Research Council under grant numbers 647112 (CH and MA) and 693024 (SJ). 
CB acknowledges the support of the Australian Research Council through the award of a Future Fellowship.
DL acknowledges support from the McWilliams Center for Cosmology, Department of Physics, Carnegie Mellon University.
HHi acknowledges support from an Emmy Noether grant (No. Hi 1495/2-1) of the Deutsche Forschungsgemeinschaft.
HHo acknowledges support from Vici grant 639.043.512, financed by the Netherlands Organisation for Scientific Research (NWO).
BJ acknowledges support by an STFC Ernest Rutherford Fellowship, grant reference ST/J004421/1.
JHD acknowledges support from the EuropeansCommission under a Marie-Sk{l}odwoska-Curie European Fellowship (EU project 656869).
SJ also acknowledges support from the Beecroft Trust.
DP acknowledges the support of the Australian Research Council through the award of a Future Fellowship.
MB is supported by the Netherlands Organisation for Scientific Research, NWO, through grant number 614.001.451.
%CM acknowledges support from the National Science Foundation through Cooperative Agreement 1258333 managed by the Association of Universities for Research in Astronomy(AURA), and the Department of Energy under Contract No. DE-AC02-76SF00515 with the SLAC National Accelerator Laboratory. 
%KK acknowledges support by the Alexander von Humboldt Foundation research award.
%EvU acknowledges support from an STFC Ernest Rutherford Research Grant, grant reference ST/L00285X/1, and BJ from an STFC Ernest Rutherford Fellowship, grant reference ST/J004421/1.
%MV acknowledges support from the European Research Council under FP7 grant number 279396 and the Netherlands Organisation for Scientific Research (NWO) through grants 614.001.103. 
%JdJ is supported by the Netherlands Organisation for Scientific Research (NWO) through grant 621.016.402.

This work is based on data products from observations made with ESO Telescopes at the La Silla Paranal Observatory under programme IDs 177.A-3016, 177.A-3017 and 177.A-3018, and on data products produced by Target/OmegaCEN, INAF-OACN, INAF-OAPD and the KiDS production team, on behalf of the KiDS consortium.

2dFLenS is based on data acquired through the Australian Astronomical 
Observatory, under program A/2014B/008.  It would not have been possible 
without the dedicated work of the staff of the AAO in the development 
and support of the 2dF-AAOmega system, and the running of the AAT.

We thank the GAMA consortium for providing access to their third data release. GAMA is a joint European-Australasian project based around a spectroscopic campaign using the Anglo-Australian Telescope. The GAMA input catalogue is based on data taken from the Sloan Digital Sky Survey and the UKIRT Infrared Deep Sky Survey. Complementary imaging of the GAMA regions is being obtained by a number of independent survey programs including GALEX MIS, VST KiDS, VISTA VIKING, WISE, Herschel-ATLAS, GMRT and ASKAP providing UV to radio coverage. GAMA is funded by the STFC (UK), the ARC (Australia), the AAO, and the participating institutions. The GAMA website is http://www.gama-survey.org/.

Computations for the $N$-body simulations were performed in part on the Orcinus supercomputer at the WestGrid HPC consortium (www.westgrid.ca),
in part on the GPC supercomputer at the SciNet HPC Consortium.
SciNet is funded by: the Canada Foundation for Innovation under the auspices of Compute Canada;
the Government of Ontario; Ontario Research Fund - Research Excellence; and the University of Toronto.

{\footnotesize {\it Author contributions:}  All authors contributed to the development and writing of this paper.   The authorship list is given in two groups: the lead authors (AA, CB, CH, DL), followed by an alphabetical group which includes those who have either made a significant contribution to the data products or to the scientific analysis.}

\bibliographystyle{mnras}
\bibliography{EG_mnras} % if your bibtex file is called example.bib

\appendix
\section{$E_G$ covariance}  \label{app}
In Section~\ref{sec:cov} we compute the covariance for $E_{\rm G}$ following equation~\ref{eqn:EGcov}, where we incorporate a covariance for the lensing measurement estimated from a mock analysis, $\hat{C}(\Upsilon_{\rm gm})$, a covariance of clustering measurement determined from a Jackknife analysis, $\hat{C}({\Upsilon_{\rm gg}})$ and the uncertainty on the RSD parameter, $\sigma_{\beta}$. In this Appendix, we investigate the contribution of each of these terms to the covariance for our final measurement. The  $E_{\rm G}$ covariance in equation~\ref{eqn:EGcov} can be written as the sum of a lensing term, $D$, a clustering term, $W$ and a $\beta$ term, $B$, as
\begin{equation}
\hat{C}(E_{\rm G})=D+W+B\, , 
\label{eqn:EGcovmod}
\end{equation}
where, for example,
\begin{equation}
D=(E_{\rm G}^iE_{\rm G}^j)\frac{\hat{C}(\Upsilon_{\rm gm})^{i,j}}{\Upsilon_{\rm gm}^i\Upsilon_{\rm gm}^j}  \, , 
\label{eqn:D}
\end{equation}
with the diagonal components denoted as $D^d$ where, 
\begin{equation}
D^d=(E_{\rm G}^i)^2\frac{\hat{C}(\Upsilon_{\rm gm})^{i=j}}{(\Upsilon_{\rm gm}^i)^2}  \, .
\end{equation}

\begin{figure*}
\centering
\begin{subfigure}[b]{\textwidth}
   \includegraphics[width=0.95\textwidth]{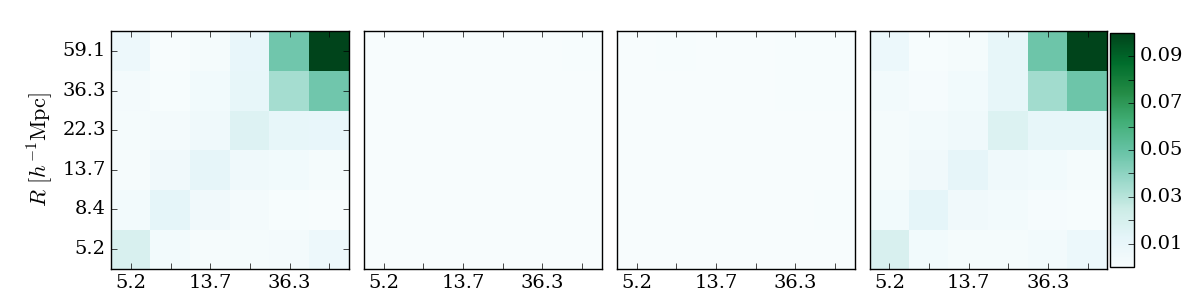}
   \caption{CMASS}
\end{subfigure}

\begin{subfigure}[b]{\textwidth}
   \includegraphics[width=0.95\textwidth]{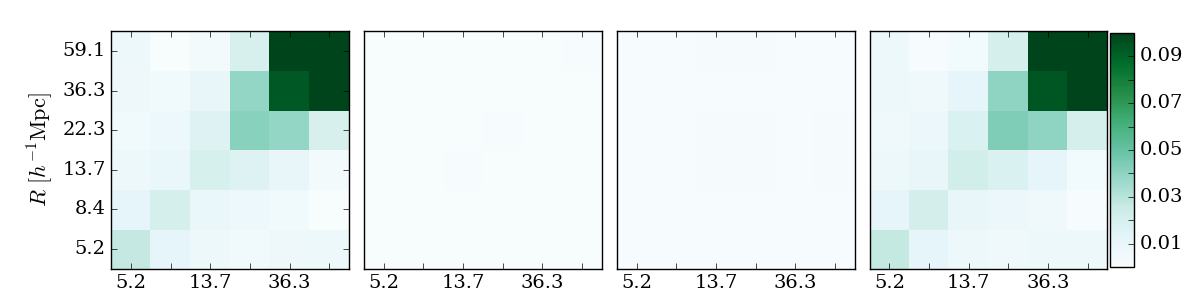}
   \caption{LOWZ}
\end{subfigure}

\begin{subfigure}[b]{\textwidth}
   \includegraphics[width=0.95\textwidth]{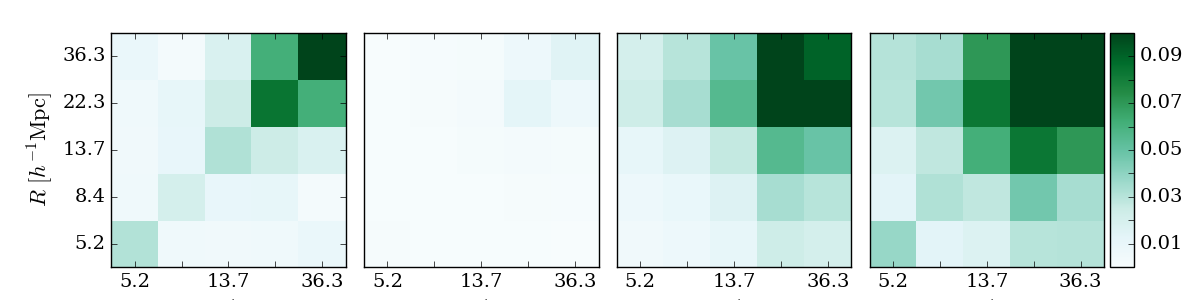}
   \caption{2dFHIZ}
\end{subfigure}

\begin{subfigure}[b]{\textwidth}
   \includegraphics[width=0.95\textwidth]{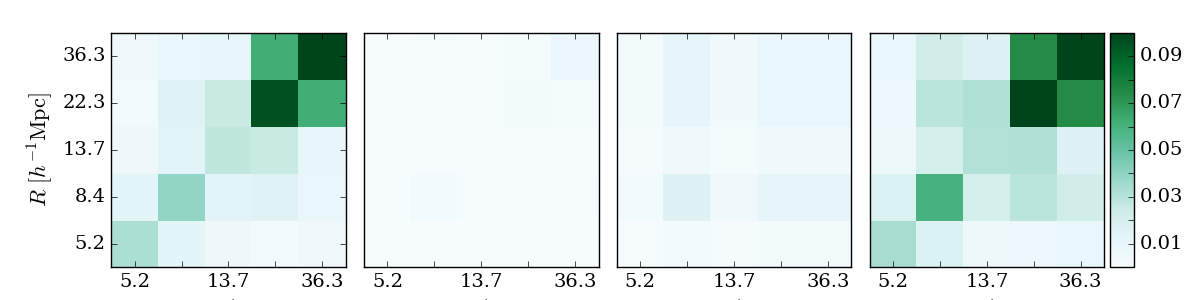}
   \caption{2dFLOZ}
\end{subfigure}

\begin{subfigure}[b]{\textwidth}
   \includegraphics[width=0.95\textwidth]{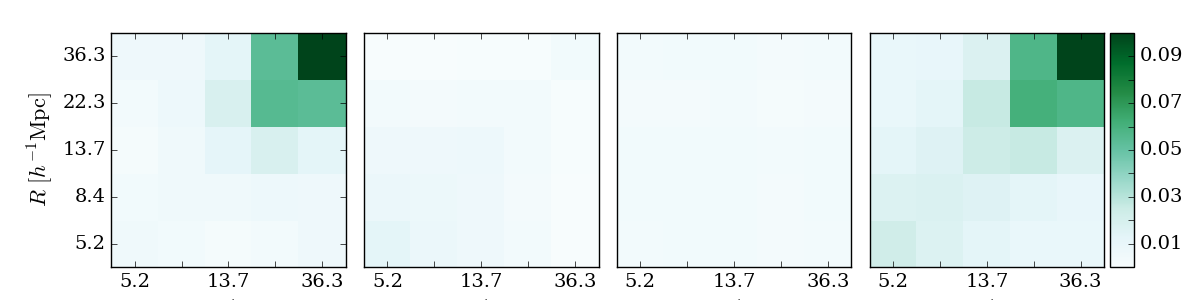}
   \caption{GAMA}
\end{subfigure}

\caption[Two numerical solutions]{ The three different components of the $E_{\rm G}(R)$ covariance in equation~\ref{eqn:EGcovmod} and~\ref{eqn:EGcov} and their sum, for the analyses with each lens sample. From left to right the panels show $D$, $W$, $B$ and $\hat{C}(E_{\rm G})$ with a consistent colour scale.}
\label{fig:CGcov}
\end{figure*}

Figure~\ref{fig:CGcov} represents the different components of $\hat{C}(E_{\rm G})$ for the analyses for each of the lens samples. In all cases it is evident that the left-hand panel, which shows the lensing covariance, $D$, defined in equation~\ref{eqn:D}, dominates compared to $W$ the clustering jack-knife covariance. This is expected as the lensing measurement is dominated by shape noise. This justifies the use of a jack-knife covariance for the clustering measurement, rather than a full mock analysis. Furthermore, for all of the lens samples except for GAMA, the galaxy-galaxy lensing measurement uses a significantly smaller area compared to that for the clustering and RSD measurement, rendering these measurements essentially independent.  For the 2dFLenS analyses, especially 2dFHIZ, the contribution to the error budget from the RSD parameter is more significant. For the case of GAMA, with equal areas for all components of the $E_{\rm G}$ measurement, we show that while the lensing covariance is dominant, $W$ and $B$ are significant.

\begin{figure*}
\centering

\begin{subfigure}[b]{\textwidth}
\includegraphics[scale=0.37]{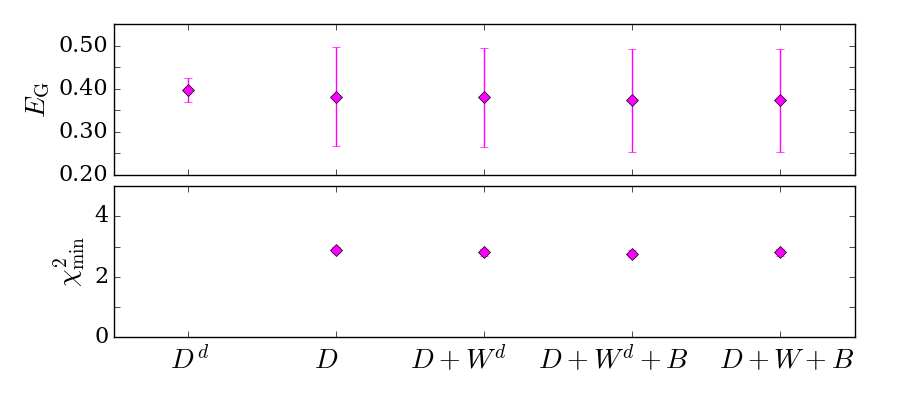}
\includegraphics[scale=0.37]{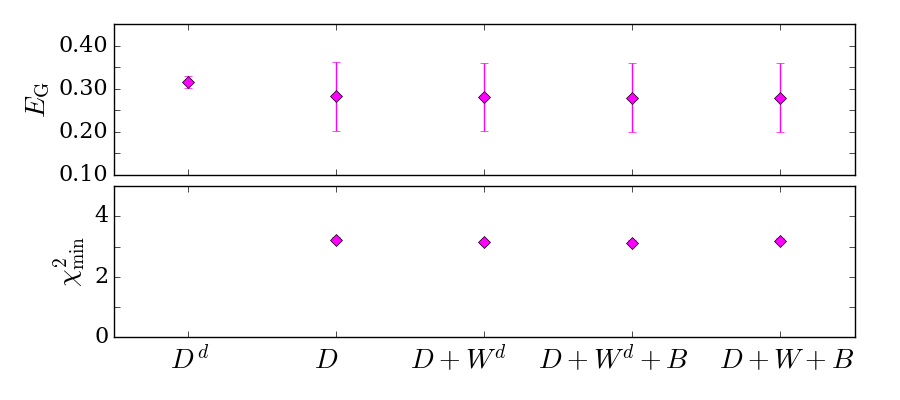}
\caption{LOWZ (left) and CMASS (right)}
   %\label{fig:Ccov} 
\end{subfigure}

\begin{subfigure}[b]{\textwidth}
\includegraphics[scale=0.37]{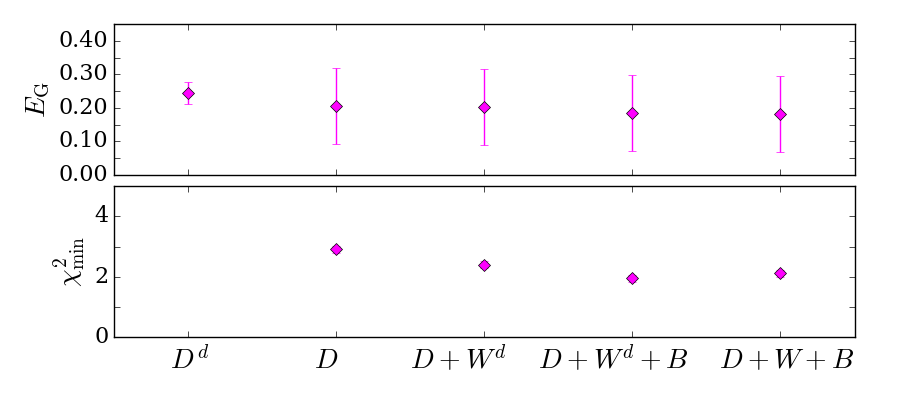}
\includegraphics[scale=0.37]{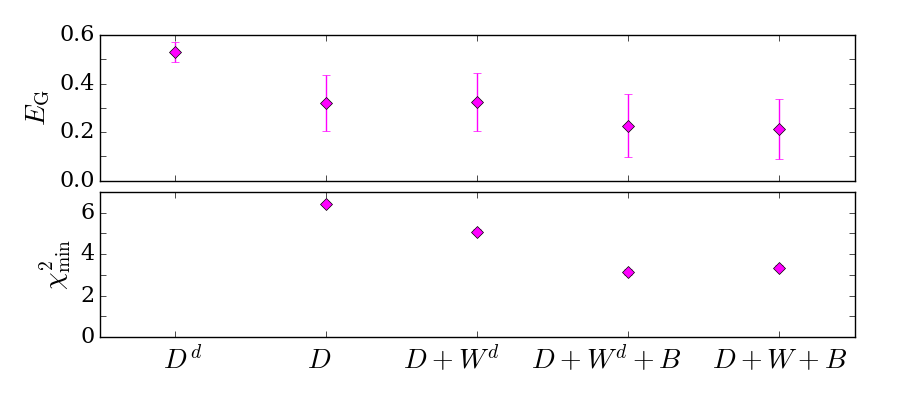}

   \caption{2dFLOZ (left) and 2dFHIZ (right)}
\end{subfigure}

\begin{subfigure}[b]{\textwidth}
\centering
\includegraphics[scale=0.37]{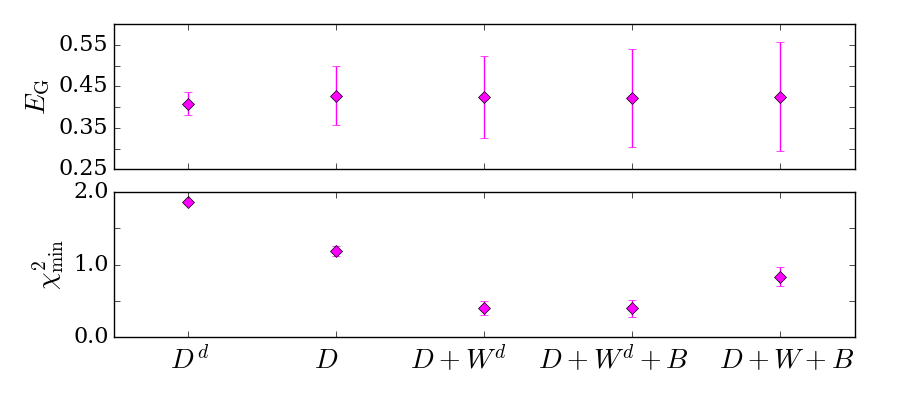}
   \caption{GAMA}
\end{subfigure}
\caption[Two numerical solutions]{ The best fit value and 1$\sigma$ uncertainty in the fit for the $E_{\rm G}(R)$ measurements (upper panels) and the associated $\chi^2_{\rm min}$ for the fit (lower panels) for the analyses with LOWZ, CMASS, 2dFLOZ, 2dFHIZ and GAMA. From left to right along the horizontal axis, different components we added to the $E_{\rm G}$ covariance (given in equation~\ref{eqn:EGcovmod}) in succession. For example, the first two data points compare the effect of using only the diagonal of the lensing covariance obtained from the mock analysis, $D^d$, with the full covariance $D$.}

\label{fig:egerrs}
\end{figure*}

To further investigate the covariance, we compare the effects of the different components of the covariance on the best fit value and 1$\sigma$ uncertainty in the fit for the $E_{\rm G}(R)$ measurement, as well as the associated $\chi^2_{\rm min}$ for the fit. The upper panels of Figure~\ref{fig:egerrs} show the best scale-independent fits to the $E_{\rm G}(R)$ measurements for each of the lens samples and how they vary as the complexity of the $E_{\rm G}$ covariance is increased. That is, we consider using the diagonal of the lensing covariance computed from the mock analysis, $D^d$ compared to the full covariance, $D$ and follow the same convention adding in the clustering and RSD components. The lower panels show the associated $\chi^2_{\rm min}$ values for these fits. 

In all cases, the most significant change in the best fit $E_{\rm G}$, the associated uncertainty and the $\chi^2_{\rm min}$ was between $D^d$ and $D$, emphasising the importance of the off-diagonals in the lensing covariance.
For CMASS, as is the case for LOWZ, the best-fit $E_{\rm G}$ and the $\chi^2_{\rm min}$ are stable to the inclusion of the clustering and beta uncertainties, though the uncertainty on the fits increase. 

For the cases of 2dFHIZ and 2dFLOZ shown in the middle panel of Figure~\ref{fig:egerrs}, the penultimate data point shows that the effect of including the uncertainty on the RSD parameter is to lower the best fit $E_{\rm G}$. As revealed in Figure~\ref{fig:CGcov}, as the relative uncertainty of the RSD parameter is large for these two lens samples, the covariance between the large scales of the $E_{\rm G}$ measurement is amplified and therefore down-weighted. Therefore, the best-fits to the data are slightly lower than expected when performing a ``chi-by-eye" analysis of Figure~\ref{fig:eg}.

For GAMA, we again show that the clustering covariance and the beta uncertainty contribute significantly to the final covariance for $E_{\rm G}$. While the best-fit does not change with increasing complexity, the $\chi^2_{\rm min}$ values do. The spuriously low $\chi^2_{\rm min}$ for the fits that include either $W^d$ or $W$ suggests that the clustering measurements are overestimated due to the size of the Jack-knife samples and this causes an over-estimation of the uncertainty on our final measurement. Furthermore, the difference between the $\chi^2_{\rm min}$ for $W^d$ and $W$ suggest that the Jack-knife analysis for the clustering overestimates the uncertainty due to the limited Jack-knife box size.

We note that we have not accounted for any covariance between the clustering and RSD measurement. This effect would be more significant for 2dFLenS and GAMA as in these cases, the uncertainty on the lensing measurement is less dominant. However, both the uncertainty on the RSD measurements and the clustering measurements are shown in Figure~\ref{fig:egerrs} to have at most, a ten percent shift on the value of $E_{\rm G}$, compared to the uncertainty on the measurement, which is roughly fifty percent. As such, we assume that any covariance between the RSD and clustering measurements will contribute less and can be safely ignored, given the precision of this analysis.

% Don't change these lines
\bsp	% typesetting comment
\label{lastpage}
\end{document}